# Mechanical self-assembly of a strain-engineered flexible layer: wrinkling, rolling, and twisting

Zi Chen,[1,*] Gaoshan Huang,[2] Ian Trase,[1] Xiaomin Han,[1] Yongfeng Mei[2,*]

[1.] Thayer School of Engineering, Dartmouth College, Hanover, NH 03755, USA

[2.] Department of Materials Science, Fudan University, Shanghai 200433, P. R. China

*Email address: zi.chen@dartmouth.edu; yfm@fudan.edu.cn. Zi Chen, Gaoshan Huang, and Ian Trase contributed equally.

## Abstract

Self-shaping of curved structures, especially those involving flexible thin layers, has attracted increasing attention because of their broad potential applications in e.g. nanoelectromechanical/micro-electromechanical systems (NEMS/MEMS), sensors, artificial skins, stretchable electronics, robotics, and drug delivery. Here, we provide an overview of recent experimental, theoretical, and computational studies on the mechanical self-assembly of strain-engineered thin layers, with an emphasis on systems in which the competition between bending and stretching energy gives rise to a variety of deformations, such as wrinkling, rolling, and twisting. We address the principle of mechanical instabilities, which is often manifested in wrinkling or multistability of strain-engineered thin layers. The principles of shape selection and transition in helical ribbons are also systematically examined. We hope that a more comprehensive understanding of the mechanical principles underlying these rich phenomena can foster the development of new techniques for manufacturing functional three-dimensional structures on demand for a broad spectrum of engineering applications.

# I. Introduction

The spontaneous bending, twisting, and wrinkling of thin layers are ubiquitous in both natural and synthetic systems. These phenomena have garnered significant interests from the scientific community because of their potential applications in sensors [1], actuators [2], micro-robotics [3], nanoelectromechanical systems (NEMS) [4], active materials [5], optoelectronics [6], stretchable electronics [7], and drug delivery systems [8]. These deformations often stem from a need to release potential energy in presence of surface stresses [9], misfit strains [10], residual strains [11-13], thermal stresses [14], swelling/shrinkage [15,16], or differential growth [17,18]. As such, deformations are often triggered by environmental factors, such as humidity, temperature, and pressure. Significant research has been devoted to causing asymmetric deformation in thin layers under uniform environmental change, often by introducing some level of anisotropy into the system. This behavior allows the construction of systems that can have multiple complex responses to simple changes in scalar signals.

In synthetic thin structures, if a flexible layer is extended with a geometrically limited boundary, wrinkling or buckling in the perpendicular direction will occur. Strain engineering in a thin layer could produce a variety of three-dimensional (3D) structures besides wrinkles. For instance, self-rolling of a layer with a strain gradient can lead to the formation of tube or scroll-like structures [2]. Experimental and theoretical investigations about formation mechanisms of these novel micro/nano-structures have previously been carried out. With the rapid development of nanostructures and nanodevices, these structures are being pursued for many applications like flexible electronics, stretchable electronics, nanophotonics, robotics, and microfluidics.

Furthermore, helices can be considered as a special subclass of structures formed through rolling. They are chiral structures with broken left-right symmetry [19], which can be either left-handed or right-handed with a mirror image that has an opposite handedness or chirality. It is natural to think that such symmetry-breaking comes from the chirality of the microscopic building units, but this is not always the case [12]. Spontaneous helical ribbons can form either under terminal loads (e.g., tensile forces or torques) [20] or in the absence of terminal loads [9,15,21-23]. Shape transitions between purely twisted ribbons (or helicoids), cylindrical helical ribbons, and tubules have been frequently observed in twist-nematic elastomers [1,24], peptides [22,25], strained multilayer composites [9,15,21-23], and nanoribbons [10,26]. These transitions are often driven by the complex interplay between molecular interactions, environmental stimuli, elastic properties, and nonlinear geometric effects.

Twisting can exist in thin layers outside of the uniform helix shape subclass, and can occur even when the driving forces are not off-axis with respect to the geometric axes of the material [12]. Twists in thin layers have an associated handedness and can even switch handedness [17]. Off-axis competition between the internal chirality of the building units of the thin layer and macroscopic forces acting on the layer can also form twists.

The article covers analysis of these afore-mentioned deformation modes such as wrinkling, rolling and twisting, along with examples and applications of each. This is followed by a synthesis of current promising research directions and future applications. Readers interested in the mechanics of thin layer deformation are encouraged to refer to some excellent recent reviews for a more comprehensive understanding of recent developments [27-30].

Recently, wrinkling in nanoparticle films has received a great deal of attention. The most common films include graphene [31-33], nanocrystals [34-37], and nanotubes [38-40]. While the following discussion avoids focusing on nanoparticle films in depth, they are an important subset of wrinkling research and the reader is encouraged to refer to those articles. It should also be noted that while the discussion of wrinkling and rolling mechanics is primarily focused on inorganic materials, polymer sheets are becoming increasingly important in science and engineering. For a discussion of the specific mechanics of polymer surfaces, we point the reader to J. Rodriguez-Hernandez's excellent review [41].

**II. Self-assembly and strain distribution in wrinkles**

**A. Wrinkling vs. rolling in a strain-engineered flexible layer**

A flexible pre-strained layer attached to a fixed boundary, adopts a 3D structure in order to minimize strain energy [42-44]. Typical deformations reported in literature include wrinkling [45,46], bending [43,44,47], folding [48-50], and ridging [51], depending on the strain gradient and magnitude. Normally, a small strain gradient produces wrinkles while a large strain gradient makes the layer bend or roll into a curved structure [52]. Real materials are more complicated, as competition between bending and stretching energy can cause transitions between wrinkling and rolling states after the flexible layer is set free in one end and fixed in the other end.

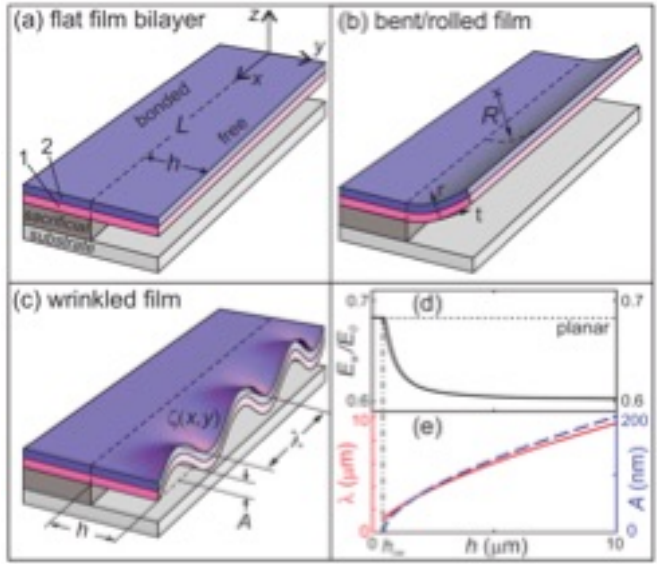

FIG. 1. Schematics of (a) released bi-layer, (b) bent bi-layer with inner radius $R$, and (c) wrinkled structure with deflection profile $\zeta(x, y)$, amplitude $A$, and wavelength $\lambda$. (d) $E_w$ (solid line) and energy of planar relaxation (dashed line) as a function of h. (e) Wavelength $\lambda$ (solid line, left axis) and amplitude $A$ (dashed line, right axis) as a function of length $h$, in the case of wrinkled structure. Vertical dotted-dashed line marks the $h_{cw}$. Reprinted with permission from Ref. [52] (Copyright 2009 by the American Physical Society).

Theoretical analyses have been carried out on both the wrinkling and rolling cases

[29,42,52-60]. A simple analysis can be done using an isotropically strained bi-layer structure. Shown in Fig. 1 [52], the bi-layer with thicknesses $d_1$ and $d_2$ is subjected to biaxial strain $\varepsilon_1$ and $\varepsilon_2$ respectively. The bi-layer is free hanging over a distance h and is initially in a strained state over the length L. The released portion is free to elastically relax, constrained only by the fixed boundary (see the dashed line in Fig. 1a) [52]. The average strain and strain gradient of the bi-layer are defined as $\bar{\varepsilon} = (\varepsilon_1 d_1 + \varepsilon_2 d_2)/(d_1 + d_2)$ and $\Delta\varepsilon = \varepsilon_1 - \varepsilon_2$ respectively, and $\Delta\varepsilon > 0$. The initial elastic energy (given per unit area) of the bi-layer is $E_0 = Y(d_1\varepsilon_1^2 + d_2\varepsilon_2^2)/(1-\nu)$, where Y and $\nu$ are Young's modulus and Poisson's ratio respectively. In the bending case, the fixed boundary limits relaxation in the x direction and the strain is relaxed via bending in the y direction to form a curved structure with inner radius R (Fig. 1b). Since the layers are thin, the stress component in the radial direction (through the thickness direction) must be zero at equilibrium [61]. The total elastic energy of the bent film $E_{bent}$ is calculated by integrating the elastic strain energy density from the outer to the inner film surface [52]. The equilibrium elastic energy of the rolled structure $E_{bent}$ can be obtained by minimizing the energy and is normalized to $E_0$ and then compared with the wrinkle energy $E_w$ [52]. In the wrinkling case (Fig. 1c), the deflection of the bi-layer can be written as a function $\zeta = Af(y)\cos(kx)$, where A is the maximum amplitude of the wrinkle at the free end, k is the wrinkle wave number in the x direction, and $f(y) = [1-\cos(\pi y/h)]/2$. In the calculation of the wrinkle energy, $E_w$ is averaged over one wavelength, L=$\lambda$, and L is numerically minimized with respect to A, $\lambda$, and $\gamma$ (where $\lambda$ is the wavelength of the wrinkle, and $\gamma$ is the magnitude of relaxation in the y direction). The wrinkle energy as a function of wrinkle length is given in Fig. 1d, and it is found that there is a minimum critical wrinkle length h$_{cw}$, for wrinkle formation. The value of h$_{cw}$ is $h_{cw} \approx 2.57 d_2 \sqrt{-\bar{\varepsilon}}$ [52]. For h<h$_{cw}$, energy minimization provides only a trivial minimum of the wrinkle energy with A=0 and $\lambda \rightarrow \infty$ [62], corresponding to a planar relaxation in the y direction (dashed line in Fig. 1d). For h>h$_{cw}$, wrinkling can occur, and both $\lambda$ and A increase with h, as demonstrated in Fig. 1e. The preferred equilibrium shape of a free-hanging film could be found by comparing these two normalized energies, E$_{bent}$ and E$_w$. h>h$_{cw}$ does not guarantee wrinkle formation. Large h also allows the bi-layer to roll if the strain gradient (or $\Delta \varepsilon$) is also large. A wrinkled structure is only formed when the strain gradient is small enough; even if h>h$_{cw}$ [29,52]. For a typical bi-layer consisting of 10 nm In$_{0.1}$Ga$_{0.9}$As and 10 nm GaAs, Young's modulus Y=80 GPa, and Poisson ratio $\upsilon$=0.31, the $\varepsilon_1$ and $\varepsilon_2$ were systematically changed to calculate the favorable shape as a function of h and $\Delta \varepsilon$, and the obtained phase diagram is shown in Fig. 2. When $\Delta \varepsilon$=0.20 % and $\bar{\varepsilon}$=-0.36 %, bending is favored when h<700 nm. With larger h, the wrinkled structure has lower energy and is favorable [52]. The boundary between the two shapes is shown as a solid line in Fig. 2. For higher $\bar{\varepsilon}$ like -1.0 %, the phase boundary curve moves upwards (see dashed line in Fig. 2) and the wrinkling region is enlarged [52]. Fig. 2 shows that for wrinkled structure, the wavelength $\lambda$ increases with h, while for bent structure, the equilibrium radius R$_{eq}$ decreases with increasing $\Delta \varepsilon$. One should note that at small length scales, continuum theory may not always be accurate, as misfit dislocations in the boundary, surface properties, and size effects play an increasing large role.

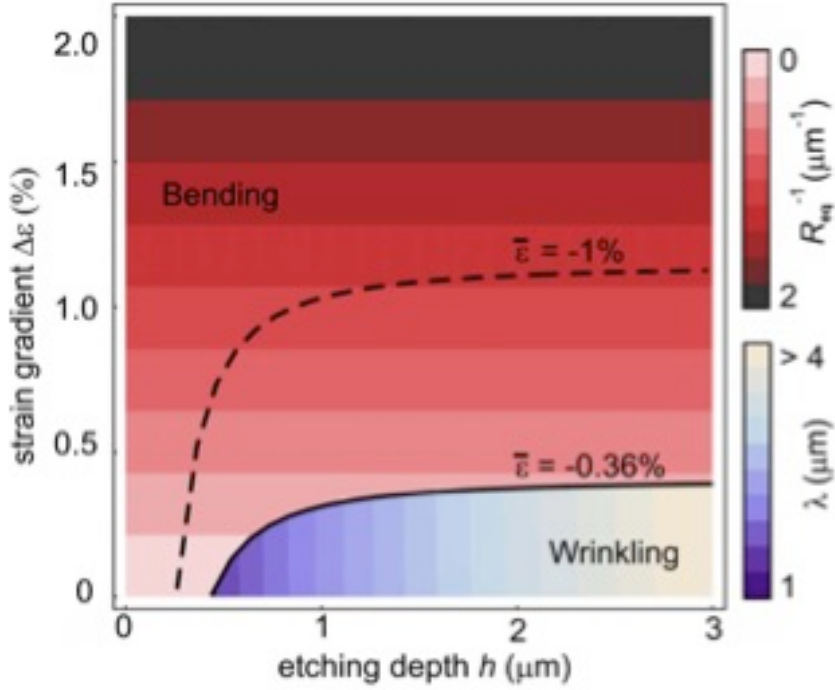

FIG. 2. Phase diagram of favorable shapes of strained bi-layer based on the energetic comparison between $E_w$ and $E_{bent}$. Solid curve indicates the boundary separating the bent and wrinkled shapes. $R_{eq}$ is shown for the bent structure and wavelength $\lambda$ for the wrinkled structure. The dashed curve is the phase boundary curve for $\bar{\varepsilon}$ =-1.0%. Reprinted with permission from Ref. [52] (Copyright 2009 by the American Physical Society).

For an anisotropically strained flexible layer, the situation is simpler. The coordinate system is the same as in Fig. 1a. If the y direction is the most compliant direction, the layer will bend or roll in this direction [29]. If the x direction is the most compliant direction, deformation is dependent on $h_{cw}$. When $h<h_{cw}$ the boundary (dashed line in Fig. 1a) does not allow the strain to relax along x direction and the strain will be retained, but when $h>h_{cw}$ the released layer attempts to bend in the x direction and the constraint thus causes the wrinkles [29,42].

In the above investigation, only pure bending and pure wrinkling are considered, and the favorable shapes are decided by the energy minimization. However, experimental results sometimes show obvious deviation from these theoretical predictions and the parameters during release can considerably influence the final geometry [63]. For example, Figures 3a-3d show the SEM images of the sample morphologies as a function of etching time, and the superposition of bending/rolling and wrinkling is obvious [64]. For an etching distance h = 3.4 μm (Fig. 3a), one can see a bent layer corresponding roughly to 1/6 of a tube circumference along with wrinkles of wavelength $\lambda$ = 30 ± 3 μm on its edge. For h = 5.7 μm, the released layer is rolled up to about 1/3 tube circumference, and the wrinkles have a wavelength of $\lambda$ = 76 ± 10 μm (Figs. 3b and 3c). The layer with the etching distance of h = 10.3 μm has rolled up to about 1/2 tube circumference with large wrinkles of wavelength $\lambda$ = 124 ± 19 μm, as shown in Fig. 3d. These results show co-existence of rolling and wrinkling and demonstrate a large departure from previous analytical scaling predictions [64]. The strained bi-layer is forced to accommodate both the rolled-up and the wrinkled morphology (due to the strain gradient and the average strain present in the bilayer) even though it is energetically the less favorable state. In such cases, finite-element method (FEM) simulation gives more accurate predictions. The anomalous results are believed to be the result of additional effects during relaxation of strained bi-layer, such as stress focusing and capillary forces, which can have a significant effect but were not taken into account in the analytical calculations [64,65].

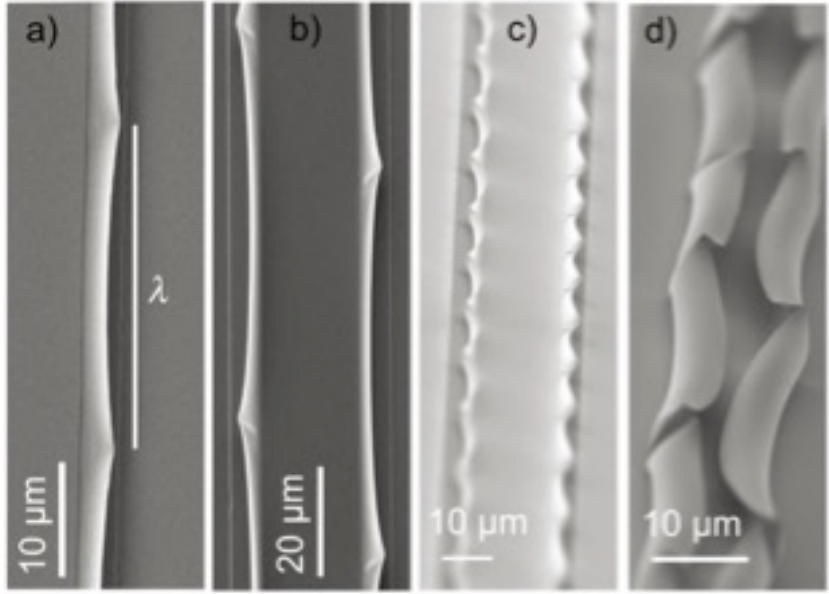

FIG. 3. Morphology evolution for increasing etching time. SEM images are shown for: (a) $h = 3.4$ μm (top view); (b) $h = 5.7$ μm (top view); (c) $h = 5.7$ μm (side view), (d) $h = 10.3$ μm (side view). Reprinted with permission from Ref. [64].

The relationship between bending and wrinkling can be more than just competition or co-existence. Researchers investigated the bending behavior after releasing an initially wrinkled layer, and found that rolling parallel to wrinkle is more favorable due to the energy barrier existing in the calculation [65]. However, the external forces exerted during the fabrication process can strongly influence the final geometry [65]. While not discussed in this review, the transition between wrinkling and crumpling is another important mechanism governed by primarily geometric constraints [66], especially in crystalline sheets [67].

**B. Internal and external control for a wrinkled layer**

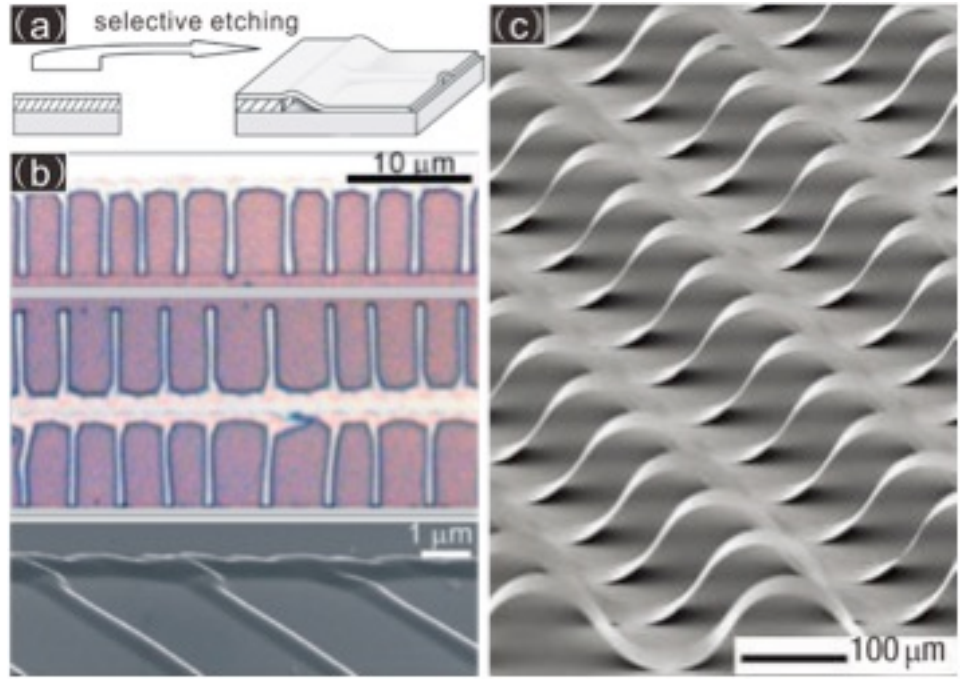

FIG. 4. (a) A schematic illustrating the formation of the wrinkles perpendicular to the etching front. The subsequent bond-back effect leads to the formation of nanochannel network. Adapted with permission from Ref. [68]. (b) Optical microscopy images showing a linear nanochannel network with single-sided (upper panel) and double-sided branch channels (middle panel). A corresponding SEM image (bird view) of a single-sided linear nanochannel network is given in the lower panel. Adapted with permission from Ref. [68]. (c) Si ribbon structures formed on a PDMS substrate pre-strained to 50%. Reprinted by permission from Macmillan Publishers Ltd: Nature Nanotechnology, Ref. [69] (Copyright 2006).

Practically, there are two methods frequently used to build and control wrinkling structures. The first approach, the internal control method, introduces internal strain into the active layer and is consistent with aforementioned theoretical model. In 2007, Mei et al. [68] reported the formation of SiGe wrinkle arrays using this method. Briefly, A thin strained functional layer (e.g. SiGe layer), deposited onto a sacrificial buffer layer is partially released from the surface of the substrate by selectively etching off the sacrificial layer. Once the strained functional layer is freed from the substrate while one side is still fixed on the substrate, the strain elastically relaxes and causes wrinkles perpendicular to the etching front. The subsequent bond-back effect to the substrate leads to the formation of the ordered nanochannel network (Fig. 4a) [68]. Fig. 4b shows the SiGe wrinkles formed with a straight etching front produced by photolithography or mechanical scratching. The network shown in the upper image of Fig. 4b consists of a single-sided branched channel network directly connected to the main channel. The middle image shows a double-sided branch nanochannel network, with the main channel running in between the wrinkled branch channels. Fig. 4b is a top-down SEM shows that the open branch channel ends are arcs 300-500nm wide and 120nm high. The geometry of the wrinkle (channels with arc-shaped cross-section) deviates from the ideal case used in the theoretical calculation and the measured periodicity is smaller than the value predicted by calculation [68]. This discrepancy is due to the interaction between the free-hanging film and the substrate once the wrinkling amplitude becomes larger than the thickness of the sacrificial layer (which was not considered in the previous model). If the wrinkled film partially bonds back to the surface of the substrate, the layer cannot adapt its equilibrium periodicity, resulting in a decrease of practical periodicity. However, the film is not expected to tightly bond back to the surface during under-etching, making periodicity increase with longer etching length. Although the wrinkles in the present case are not ideally sinusoidal due to the bond-back effect, the channels formed may have important applications in micro-/nano-fluidics and biology [68]. Since the wrinkles are perpendicular to the etching front and the etching front can be well defined by conventional photolithography, complex channel arrays can be produced [70]. Malachias et al. [70] found that in two-dimensional (2D) case, the self-assembly of the channels is influenced by shape, size, spacing of the etching start windows, and layer thickness. Under optimal conditions, ordered micro-/nanochannel arrays could be formed.

By contrast, the second approach, the external control method, does not require the release of the functional layer from the substrate. In this approach, the substrate is compressed or shrunken, and the surface functional layer is "expanded" related to the substrate, causing wrinkles.[69,71] A typical wrinkle structure produced using this approach is shown in Fig. 4c. To produce ordered wrinkle arrays, Sun et al. [69] patterned surface chemical adhesion sites on pre-strained polydimethylsiloxane (PDMS) substrate by UV light illumination through a photomask. Exposure to UV light creates patterned areas of ozone proximal to the surface of the PDMS [72]. The ozone converts the unmodified hydrophobic surface to a highly polar and reactive surface (activated surface), which allow various inorganic surfaces to form strong chemical bonds [73]. The unexposed sections therefore interact only weakly with other surfaces [74]. The inorganic ribbons (here: Si) were transferred to the treated PDMS substrate after baking in an oven at 90$^{\circ}$C for 5 min. Heating facilitated conformal contact and the formation of strong siloxane linkages between the activated areas of the PDMS and the native SiO$_2$ layer on the Si ribbons. Relaxing the strain in the PDMS led to the formation of Si wrinkles through the physical separation of the ribbons from the inactivated regions of the PDMS [69]. The geometry of the wrinkles produced is tunable by changing the size and distribution of the activated regions. Inorganic semiconductor wrinkles have important applications in flexible electronics and stretchable electronics shown by the works from the Rogers group at the University of Illinois at Urbana-Champaign [75-77].

It is worth pointing out that the external control method is not limited to relaxing pre-strained substrates [78,79]. Processes which produce substrate deformation may also be used. Bowden et al. [80] found that wrinkles in thin metal films formed due to thermal contraction of an underlying substrate. In fact, wrinkles can be easily produced on a treated PDMS substrate. Chung et al. [81] reported plasma-assisted wrinkle formation. The process includes: (i) heating the PDMS; (ii) exposing the PDMS to oxygen plasma to obtain a thin film of silica-like materials; (iii) cooling down to generate the wrinkles. The wavelength and amplitude of wrinkles can be controlled by heating temperature and time of plasma treatment. Generally, a higher temperature induces larger strain, leading to larger wrinkle amplitude. Increased exposure time thickens the silica-like layers and affects both the wavelength and amplitude [81].

**C. Strain distribution of wrinkled layers: neutral plane**

A pre-strained flexible layer can form wrinkled structures when minimizing its elastic energy [52,54,59,64,68]. But it is worth noting that final wrinkled structure is not strain-free. Due to the curved flexible layer, the strain has a distribution profile along the depth. This distribution can be simulated by FEM or analytically modeled by considering the geometry before and after wrinkling [82,83]. In this section, we will specifically discuss experimental results about strain distribution in wrinkles. The position of the neutral plane (where strain is zero) is a key point in this part.

In a simplified model [84,85] with a curved layer of thickness $t$ and curvature $r$, the peak strain is given by

$$\varepsilon = \frac{t}{2r} \times 100\%. \qquad (1)$$

The strain is compressive on one surface and tensile on the other, with an approximately linear variation between the two extremes and the neutral plane in the middle [85]. The location of the neutral plane is crucial for designing flexible electronics, since the device layer should be there to avoid destructive bending [86]. For a device fabricated on the surface of a substrate (e.g. PDMS), the addition of a compensating layer on top is necessary to adjust the location of the neutral plane. Kim et al. [86] used the method to calculate the location of the neutral plane in a multi-layer system. For layers from 1 to n with strain moduli of $E_1,...E_n$ and thicknesses of $t_1,...t_n$, the neutral plane can be characterized by a distance $d$ from the top surface and $d$ is given by

$$d = \frac{\sum_{i=1}^{n} E_i t_i [(\sum_{j=1}^{i} t_j) - \frac{t_i}{2}]}{\sum_{i=1}^{n} E_i t_i}. \qquad (2)$$

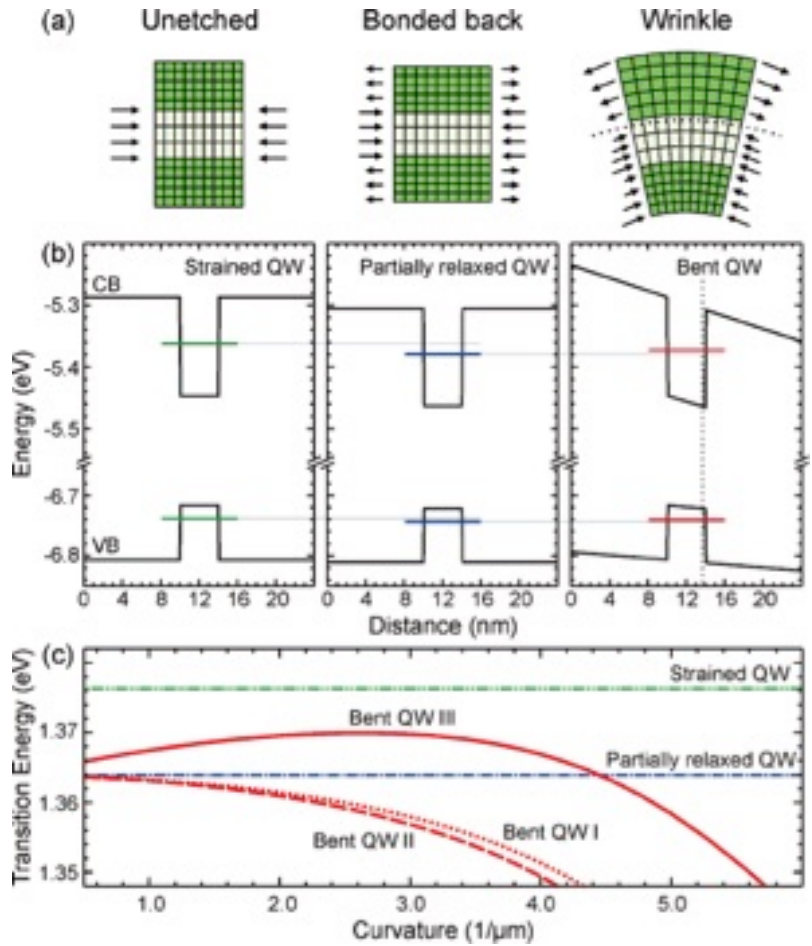

FIG. 5. (a) Schematic of the unetched (strained QW), bonded back (partially relaxed QW), and wrinkled (bent QW) structures. (b) Band diagram and quantized energy levels corresponding to the structure shown in (a). Light-gray lines are guides to the eyes. (c) Transition energy of the bent QW as a function of the bending curvature. Three bent QW models are presented. Reprinted with permission from Ref. [87] (Copyright 2007 American Chemical Society).

However, the real geometry of the 3D structure may deviate from theoretical simulation (e.g. the bond-back effect [68]), making calculations not always accurate. In addition, the experimental evidence of the strain distribution can rarely be directly obtained, especially in a very thin flexible layer. Optical methods are an effective way to probe the strain status in materials, although they are indirect methods that then need further fitting. Since the strain modifies the band structure of materials, the light emission from a wrinkled light-emitting layer can change. Luminescent InGaAs quantum wells (QW) were embedded into a GaAs nanomembrane and acted both as a local strain sensor and strain source, and therefore the shift of QW emission indicates the strain state in the layer after formation of wrinkle [87]. After the mechanical relaxation, different emission wavelengths from different regions can be detected: the emission from the under-etched region is strongly red-shifted with respect to the unetched region, and the bonded back region exhibits the longest wavelength [87]. The lattice deformations and strain states in an unetched layer and in a wrinkled layer, including the bonded-back and the wrinkled regions, are schematically displayed in Fig. 5a. The as-grown InGaAs QW has a thickness below the critical value for dislocation introduction, and therefore the compressive strain is fully confined in the QW layer before etching (left panel of Fig. 5a). After etching the layer, the bonded back region was released from the substrate and partially relaxes its internal strain. The equilibrium configuration is between the fully strained QW (without tensile strain in the barrier layers) and the fully relaxed QW (with high tensile strain in the barrier layers), and tensile strain 0.237% exists in the barrier layers (above and below the QW layer) due to strain energy minimization (see middle panel of Fig. 5a) [47,87]. For the wrinkled case in the right panel of Fig. 5a, the bending of the nanomembrane generates an inhomogeneous strain distribution where the lattice constant in the growth direction depends on the inner lattice constant and the curvature, and the strain distribution varies linearly with the distance from the inner wrinkle surface to the outer surface [87]. Assuming a position for the neutral plane (dotted line, indicating no change of the strain state before and after bending) [61], the residual forces can be obtained, which can be tensile or compressive as shown in the right panel of Fig. 5a. Based on the strain status and linear deformation potential theory [88], the band diagram and the energy levels of the QW can be obtained, as shown in Fig. 5b. The transition energies are 1.3762, 1.3638, and 1.3683 eV for strained, partially-relaxed, and wrinkled regions, respectively [87]. In order to gain more insight into the strain status of the wrinkle layer, the dependence of the calculated transition energies on the curvature are shown in Fig. 5c for three different models. The first model assumes strain energy minimization at a fixed curvature (Bent QW I). The second model assumes that the neutral plane is sitting at the center of the QW (Bent QW II). In contrast, a variable position of the neutral plane is considered as a fitting parameter in the third model (Bent QW III). Fig. 5c shows transition energies for strained and partially relaxed QWs. The first two models indicate only a redshift of transition energy compared to that of partially relaxed region, and therefore are inconsistent with the experimental results. Only the third model where the position of the neutral plane is varied gives consistent results in the curvature range of 1-2 $\mu m^{-1}$. Fig. 5c indicates that the wrinkles result in a complicated strain state, which is possibly due to the external forces originating from the bond-back effect during the drying process [47,87].

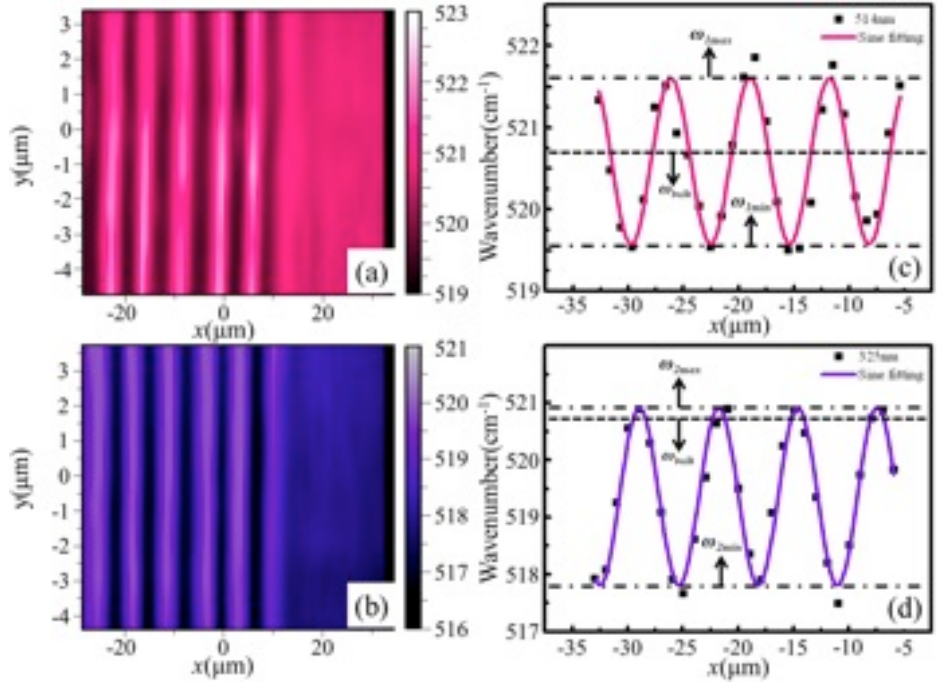

FIG. 6. Micro-Raman mapping measurements of the wrinkled Si layer, and the corresponding section analysis: (a) and (c) 514 nm laser; (b) and (d) 325nm laser. Reprinted with permission from Ref. [71] (Copyright 2013, AIP Publishing LLC).

Since Raman scattering is very sensitive to the internal strain of the flexible layer, Guo et al. [71] also carried out Raman mapping experiments on a wrinkled Si layer (on PDMS substrate) to investigate this complex strain distribution. To probe the strain distribution along the depth, two lasers (514 and 325 nm) were used as Raman excitation sources. Typical color-coded Raman peak (Si-Si TO phonon) positions extracted from mappings using the two excitation sources are shown in Figs. 6a and 6b, respectively. The peak position, i.e., the strain status at the point probed, shows periodic alternation along the x direction while keeping constant along the y direction. After the two images are aligned, it is striking to observe that the locations of the crest lines (i.e., the maximum wavenumber) of Raman mapping measured by 514 nm laser correspond exactly to the valley lines (i.e., the minimum wavenumber) of the counterpart obtained by 325 nm laser [71]. In order to simplify further investigation, a line-cut along x direction was performed, as shown in Figures 6c and 6d. The experimental results present the typical sinusoidal shape, consistent with a computed fit to a sine function, shown by the solid line. The periodicity of the peak positions along the x direction is similar to that obtained from morphological characterization [71]. For the results obtained with 514 nm laser, the Raman peak of unstrained bulk Si $\omega_{bulk}$=520.7 cm$^{-1}$ is found in the middle of the sine curve (dotted line in Fig. 6c). In 325 nm laser case (Fig. 6d), the maximum wavenumber ($\omega_{2max}$) decreases significantly and approaches $\omega_{bulk}$, and the minimum wavenumber ($\omega_{2min}$) becomes much smaller than $\omega_{bulk}$. Since the detection depth of a Raman signal correlates with the wavelength of the excitation laser, it is believed that the above phenomenon results from the inhomogeneous strain distribution along the depth of the Si layer [71].

To simplify theoretical analysis of the strain distribution, the Si layer/PDMS substrate system is separated into two basic elements: the suspended Si wrinkle and the unattached elastomeric PDMS substrate in wrinkled shape. For the suspended Si wrinkle, Von Karman elastic nonlinear plate theory [61,84,89] is used, so the dominant strain distribution in the deformed wrinkles is

$$\varepsilon_{xx}(x,y,z) = \varepsilon_{xx}^0 + \frac{\partial u_x}{\partial x} + \frac{1}{2}(\frac{\partial \omega_x}{\partial x})^2 - \frac{1}{2}\frac{\partial^2 \omega_x}{\partial x^2} z, \qquad (3)$$

where $\varepsilon_{xx}^0$ is the initial strain (the compressive strain due to the relaxation of pre-strain in PDMS, ~4.87 % for the present case), $u_x$ denotes the in-plane displacements in the x direction, $w_x$ is the deflection of the Si layer which can be written as $w_x = A\cos(kx)$, $k = \dfrac{2\pi}{\lambda}$, where $A$ and $\lambda$ are the amplitude and the wavelength of the wrinkle, respectively, and z is the distance to the center of the wrinkles. The uniform nature of strains suggests that the in-plane displacement should be written as [90]

$$u_x = \frac{1}{8} kA^2 \sin(2kx). \qquad (4)$$

Then the strain field can be written as

$$\varepsilon_{xx}(x,y,z) = \varepsilon_{xx}^0 + \frac{1}{4} k^2 A^2 + \frac{1}{2} Ak^2 \cos(kx) z . \qquad (5)$$

The strain distribution is tuned by cos(kx) and z. At the crest or the valley of the wrinkle, cos(kx) is equal to +1 or -1 and the depth distribution of the strain varies as z changes. The PDMS substrate also plays a significant role in deformation. Shearing forces at the Si layer/PDMS interface cause tension at the valley and compression at the crest, resulting in displacement of the neutral plane. Fig. 7 shows the forces in the crest and valley of the sheet, with tensile strain on the top and compressive strain on the bottom. The neutral plane, where $\varepsilon_{xx}(x, y, z)=0$, is negative at the crest and positive at the valley [71].

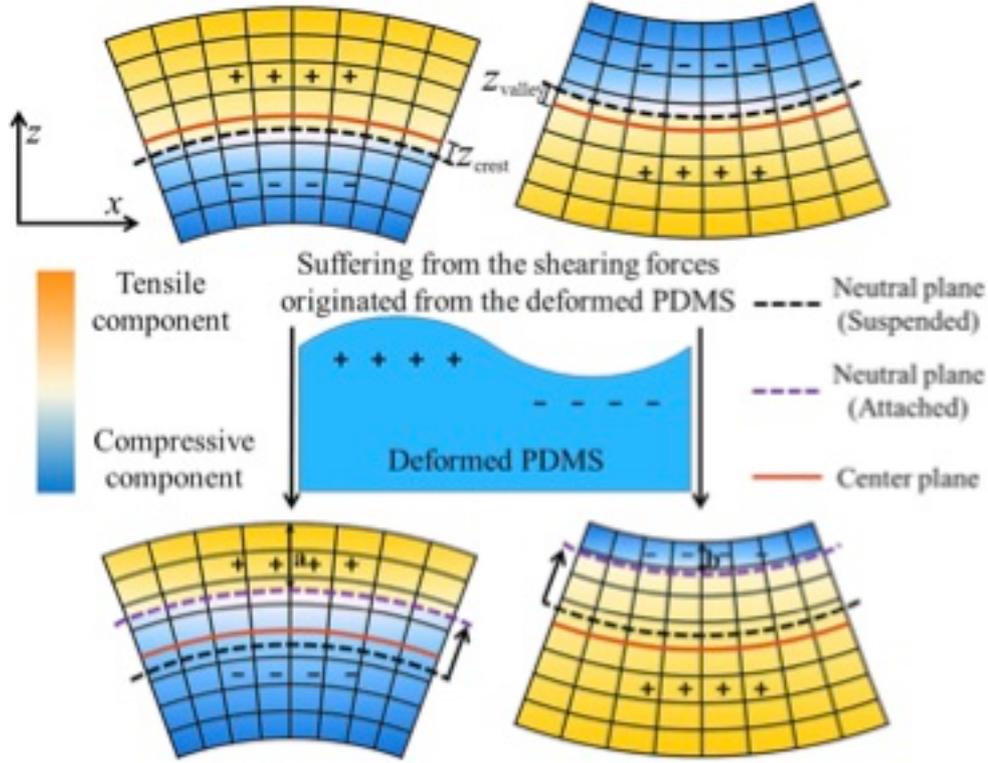

FIG. 7. Theoretical model for the strain distribution in the wrinkled silicon layer. Reprinted with permission from Ref. [71] (Copyright 2013, AIP Publishing LLC).

The detection depth of the Raman system is determined by the wavelength of the excitation source [91]. A 325nm laser can only penetrate 8nm (1/3 of the total thickness) in Si, but a 512nm laser can fully penetrate the sample. Crests generate a high wavenumber TO mode, while valleys generate a low one. The strain is also asymmetric, with the magnitude of compression at the crest higher than tension in the valley (Fig. 6c). This phenomenon is exacerbated when using a 325nm laser, as the neutral plane of the crest is similar to the maximum penetration depth. The wavenumber at the crest ($\omega_{2min}$) becomes much lower than it should be, while the valley wavenumber ($\omega_{2max}$) stays close to $\omega_{bulk}$ (Fig. 6d) [71].

## III. Fabrication and modeling of self-rolled-up tubes

### A. Mechanical self-assembly of rolled-up tubes: materials and fabrication

Strain gradients in a flexible layer can cause bending along the direction with the smallest Young's modulus, through energy minimization [92,93]. Specifically, as shown in Fig. 8, a strained bi-layer was produced with its top layer in-plane compressed and its bottom layer tensilely-stressed. When the sacrificial layer is selectively etched away, the strained bi-layer becomes detached from the substrate. Its top layer contracts, while its bottom layer expands, which results in rolling [94]. If the shape of the layer is adapted to this rolling direction, micro-/nanotubes can be formed as shown in Fig. 8 [44,94,95]. In previous literature, several approaches have been proposed to introduce the necessary strain gradient or strain difference into flexible layers with a bi- or multi-layered structure, which will be discussed in the following.

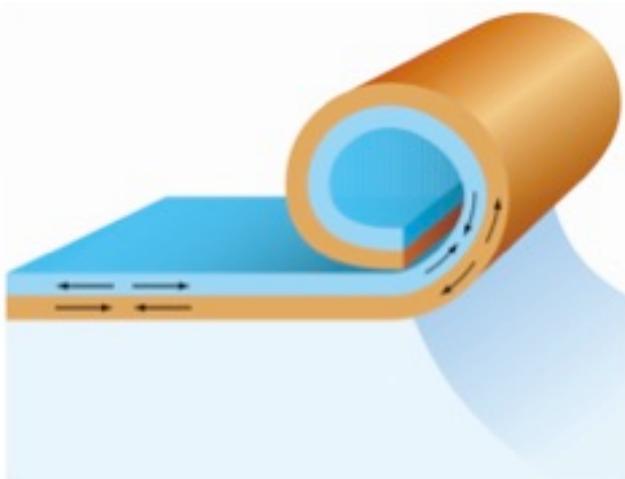

FIG. 8. When a bi-layer is freed, its top layer contracts, and its bottom layer expands, causing the bi-layer to roll-up. From Ref. [96], reprinted with permission from AAAS.

The first method utilizes lattice mismatch in an epitaxial bi-/multi-layer structure. The advantage of this method is that the strain gradient can be well controlled and thus the tube diameter is tunable (see discussion later). If the thickness of the layer is smaller than the critical value, the layer is coherent, and one can easily calculate the strain gradient based on the well-known values of the lattice constants [97,98]. For instance, a Ge layer epitaxially grown onto a Si(001) substrate creates a 4% misfit strain at the interface if the Ge is fully strained [29]. In compound semiconductor materials, the lattice constant varies with the material composition and thus allows strain tuning. Not only group IV [44,95,99,100], but also group III-V [101-103], and even II-VI [104] semiconductor materials have been strain-engineered and rolled into a tubular geometry. However, in thick layers, this method is inaccurate because the layer can be plastically relaxed via dislocations and the strain gradient cannot be well calculated by considering only lattice mismatch.

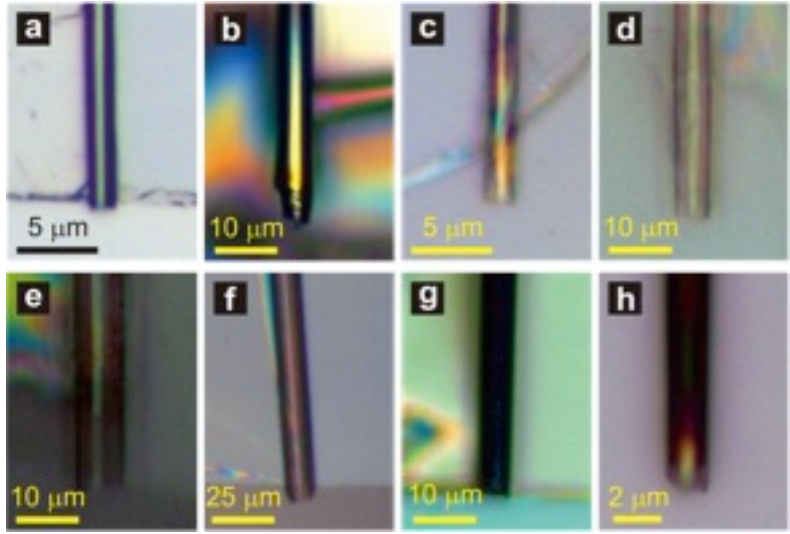

FIG. 9. Optical images of rolled-up microtubes made out of a) Pt, b) Pd/Fe/ Pd, c) TiO$_2$, d) ZnO, e) Al$_2$O$_3$, f) Si$_x$N$_y$, g) Si$_x$N$_y$/Ag, and h) diamond-like carbon. Adapted with permission from Ref. [94].

Other methods may also be used to introduce a strain gradient, but quantitative determination of the strain is difficult. The strain normally depends on experimental parameters during layer growth/deposition which can be manipulated to generate the required gradient [105]. For instance, non-epitaxial vapor deposition of thin layers may also be inherently strained [29]. In addition, it was found that the heating or cooling of materials with different thermal expansion coefficients can introduce strain into the layers [29]. By changing the experimental conditions, Mei et al. [94] have prepared microtubes from flexible layers. As shown in Fig. 9, pre-stressed inorganic layers deposited onto polymer sacrificial layers (here: photoresist) are released by removing the sacrificial layer with acetone or other organic solvents, and roll up into microtubes. Since the sacrificial layer is organic and can be easily removed by organic solvents, flexible layers of almost any inorganic material can be rolled. Figs. 9a-h show examples of rolled-up microtubes: a) Pt, b) Pd/Fe/Pd, b) TiO$_2$, b) ZnO, e)

Al$_2$O$_3$, f) Si$_x$N$_y$, g) Si$_x$N$_y$/Ag, and h) diamond-like carbon [94]. Moreover, the photoresist sacrificial layer can be easily patterned using conventional photolithography, and the flexible layer is replicated from the pattern during deposition, which is convenient for future device fabrication. Here, the layers were deposited by non-epitaxial methods, and it is believed that the strain gradient was introduced by a combination of substrate temperature evolution, deposition rate, and base pressure during deposition [94]. In multi-crystal layers, different grain sizes in the flexible layer exert different strain levels [106].

It is worth noting that although they are not commonly used, there are other approaches reported to introduce a strain gradient into a flexible layer. Theoretical investigation proved that surface reconstruction can create a strain gradient: a (2×1)-type reconstruction of Si (001) surface can create a self-driving force that bends the Si layer [107]. In polymers, distinct swelling properties of chemically dissimilar polymers in solvents can provide the strain gradient needed for bending [108]. For instance, polystyrene (PS) and poly(4-vinylpyridine) (P4VP) can be dip-coated on Si substrate to form a bi-layer structure. Upon exposure to water, PS demonstrates minimal water uptake, forming a stiff hydrophobic layer. P4VP is relatively less hydrophobic and will swell in acidic aqueous solutions because of protonation of polymer chains [108,109]. This mechanical effect was successfully employed for rolling microtubes [110].

**B. Mechanical self-assembly of rolled-up tubes: modeling**

The rolling of a flexible layer is a mechanical self-assembly process. To construct a complex 3D structure, the rolling direction and its misalignment with the crystal orientation of the flexible layer should be carefully considered. If the strain gradient was fixed, the mechanical properties of the layer would influence the rolling direction significantly. In a single-crystal epitaxial layer, anisotropy in the crystal structure leads to anisotropic mechanical properties [58]. For instance, the Young's moduli in the GaAs <100> and <110> directions are 85.3 and 121.3 GPa respectively [111,112]. The Young's moduli in Si along the <100> and <110> directions are 130.2 and 168.9 GPa respectively [113]. Calculation by energy minimization shows that rolling along the direction with the smallest Young's modulus (the softest direction) is preferred [58]. To demonstrate this phenomenon experimentally, Chun et al. [114] designed a wheel pattern with eight anchored stripe pads orientated symmetrically in the <100> and <110> directions (Fig. 10a). The results of the $In_{0.3}Ga_{0.7}As$/GaAs bilayers released from the (001) GaAs substrate are shown in Fig. 10b. The center image shows all pads around the wheel and the zoomed-in image for each pad is laid out in the outer periphery. The four pads with longer edges oriented along the <100> directions (diagonal lines of the wheel) formed tubes, with rolling taking place in the <100> direction and stopping at the foot anchors. For those four oriented along the <110> directions, the rolling still occurred in the <100> directions, and thus formed 'turn-over' triangular patterns. The persistent rolling along the <100> direction, regardless of how the rectangle stripe patterns are oriented, proves the anisotropy of stiffness in cubic GaAs crystals [114]. On the other hand, if the layer is fabricated by a non-epitaxial method, there is no anisotropy from crystal structure, but it was found that the softest direction is that perpendicular to the deposition direction [115].

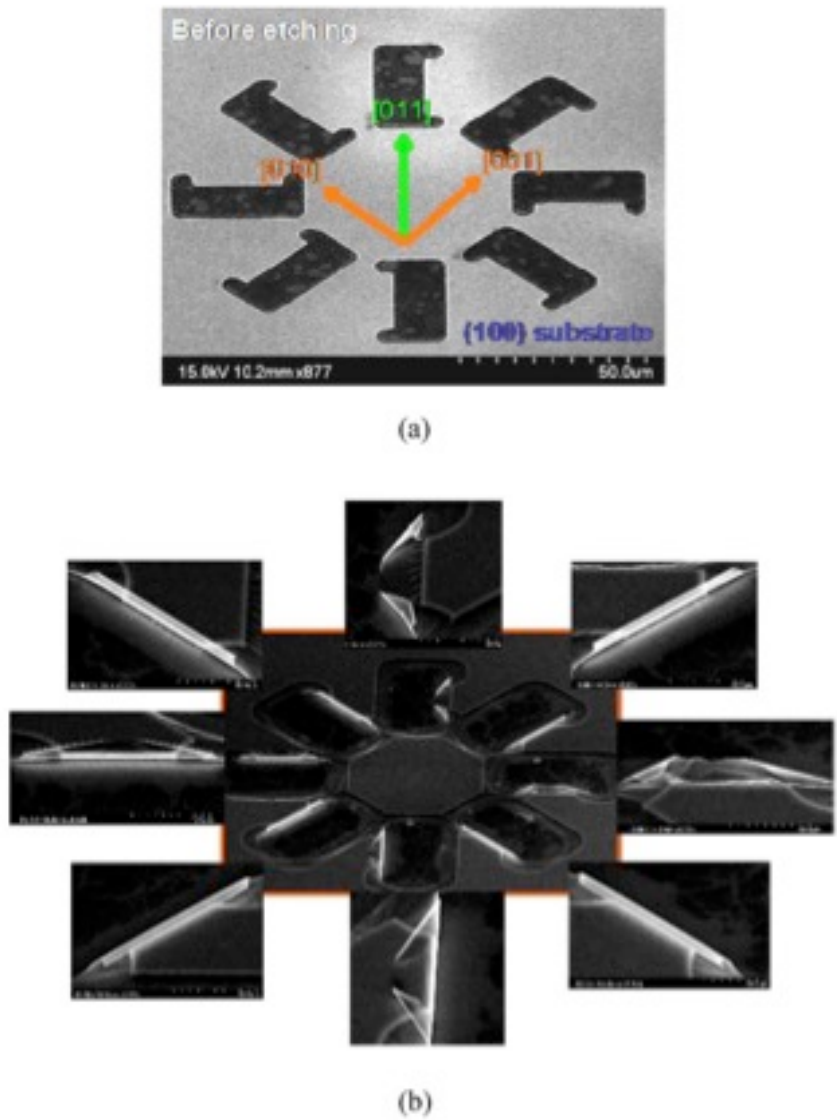

FIG. 10. A patterned wheel of anchored rectangular pads (a) before and (b) after lithography in SEM, with magnified pads displayed in (b). $In_{0.3}Ga_{0.7}As$/GaAs film was grown on (100) GaAs. Copyright 2008 IEEE. Reprinted with permission from Ref. [114].

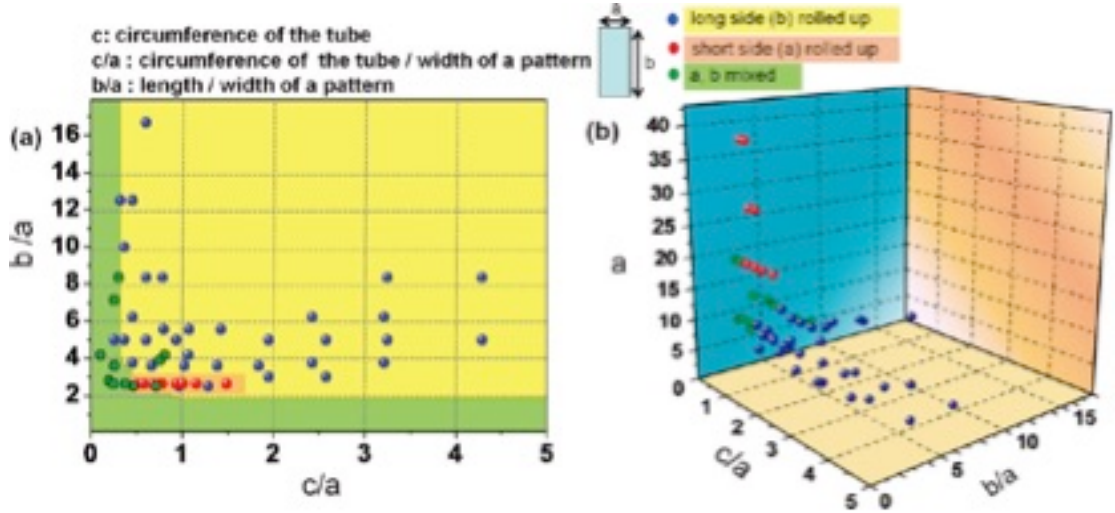

FIG. 11. Two and three-dimensional plots of rolling direction as a function of tube circumference (*c*), width (*a*), and rectangular layer length (*b*). Blue indicates long side rolling, red short side, and green for a combination. Reprinted (adapted) with permission from Ref. [63] (Copyright 2010 American Chemical Society).

Chun et al. [63] specifically investigated the rolling behavior of rectangular layer, where both sides are the softest directions, and the results are summarized in Fig. 11. The rolling direction is plotted as a function of three parameters, *b/a*, *c/a*, and *a*, where *a* is the length of the rectangular membrane, *b* is the width, and *c* is the circumference of the tube. There are three types of rolling behavior can be observed in the experiment. The blue and red dots represent long-side rolling and short-side rolling, respectively, and the green dots are mixed cases with some tubes rolled up from the long side and some from the short side. Fig. 11 shows that the rolling direction depends on not only the dimension of the starting layer (*a* and *b*) but also the tube circumference (*c*). Rolling occurs exclusively on the long side when (*c/a* > 2) or (*b/a* > 9), but is mixed when (*c/a*<<1) and (*b/a*) is low. Mixed and short rolling are also more likely to occur in sheets of larger size [63].

In an experimental study, researchers noticed that chemical etching anisotropy may also impact the rolling behavior [58]. Time evolution of rolling shows a complicated process [63]. The final rolling direction depends on the length and width of the layer, the energy of the final state, the history of the rolling process, the kinetic control of the etching isotropy, and the diameter of the tube [63].

The modeling of the tube geometry is another important concern. Considering the spiral cross-section of rolled-up tube, one can calculate the number of rotations N as [116]

$$N = -(r/d - 1/2) + \sqrt{(r/d - 1/2)^2 + L/(\pi d)}, \quad (6)$$

Where $L = \sum_{n=1}^{N} 2\pi(r + (n-1)d)$ is the rolling distance, d is the layer thickness, and r is the tube radius of the initial turn [116]. The initial diameter/radius, on the other hand, should be determined by the strain gradient, the hardness, and the thickness of the layer. A few models have been proposed in previous literature. For a bi-layer structure, the diameter D can be calculated based on a macroscopic continuous mechanical model [103,117] using the equation:

$$D = \frac{d[3(1+m)^2 + (1+m \cdot n) \cdot [m^2 + (m \cdot n)^{-1}]]}{3\varepsilon(1+m)^2}, \quad (7)$$

where $d = d_1 + d_2$ is the total thickness of the bi-layer, $\varepsilon$ is the in-plane bi-axial strain between the two layers, $n = Y_1/Y_2$ is the ratio of Young's modules, and $m = d_1/d_2$ is the ratio of the thicknesses of the two layers. In 2006, Songmuang et al. [118] proposed a slightly modified model based on their tube which rolled from partially strain-relaxed single-material layer. In such a case, the layer was divided into two regions: the lower region close to sacrificial layer experiences either a tensile or a compressive strain, and the upper layer is relaxed. The diameter of the tube can be described by [55,118]

$$D = \frac{1}{3}\frac{1}{\varepsilon}\frac{(d_1+d_2)^3}{d_1 d_2} \qquad (8)$$

It should be mentioned that the Poisson's ratio may also be considered if the material has anisotropic mechanical properties [118]. Investigations showed that calculated diameters from these models agree well with the experimental results [103,119,120], although FEM may give even more accurate predictions. The imperfection of the above models are possibly due to thickness deviation or additional strain [58]. The strain relaxation along the tube axis, as probed by x-ray diffraction [121] and optical characterization [122], may also influence the rolling process and tube geometry.

### IV. Mechanics and geometry in self-assembled helical structures

#### A. Self-assembly of amphiphilic aggregation

Chirality, or handedness, is of key importance in many physical and chemical systems. Molecular interactions, for example, are often strongly dependent on the chirality of the constituent molecules [123]. Helices are typical examples of chiral structures [19]. The self-assembly of helical structures, such as amphiphilic aggregates, is ubiquitous in natural and engineering systems, and has since served as an efficient, "bottom-up" way of manufacturing nanostructures. For example, lipid bilayers can self-assemble into helices driven by $Ca^{2+}$-mediated intermembrane binding [124]. Helical ribbons that arise through packing of amphiphilic molecules have since been investigated through both experiments and theoretical modeling [123,125-130].

Many molecules self-assemble in aqueous environments into larger aggregates, which can exhibit a variety of geometric shapes, from vesicles to twisted ribbons. Based on continuum theory, Helfrich and Prost developed a theoretical model for the bending of anisotropic membranes to interpret the formation of tubes and helical ribbons from different amphiphiles [131]. By assuming that the chiral molecules are packed with some twist with respect to the nearest neighbours, the theory showed that this molecular twist can be propagated throughout the membrane, thus creating a bending force that results in the formation of helical ribbons or tubules. Later, Ou-Yang and Liu [132] followed this membrane elasticity approach, but introduced a new linear term by viewing chiral lipid bilayers as cholesteric liquid crystals to study the helical structures. The analysis showed that a twisted ribbon shape and a helix with a $45°$ would form as intermediate states before transitioning into tubes, consistent with previous experimental observations [132]. However, more recent experiments demonstrated that helical ribbons can be stable states and that the helix angle was not necessarily $45°$ [133].

Chung et al. [126] made perhaps the first attempt to interpret the appearance of different helix angles through a theoretical approach. Subsequently, Selinger and Schnur further developed a theoretical model, based on Helfrich and Prost's continuum theory, to find that the molecular chirality or tilt can result in the helical winding of a membrane and the radius can be predicted in terms of the continuum parameters [134] (Fig. 12). Soon afterwards,

Selinger and co-workers extended this theory and considered membrane anisotropy to interpret the possibility of modulated state of tubules (Fig. 12c). Yet another contribution from Selinger et al. [129] was to show that helical ribbons can actually be equilibrium configurations instead of being just intermediate states.

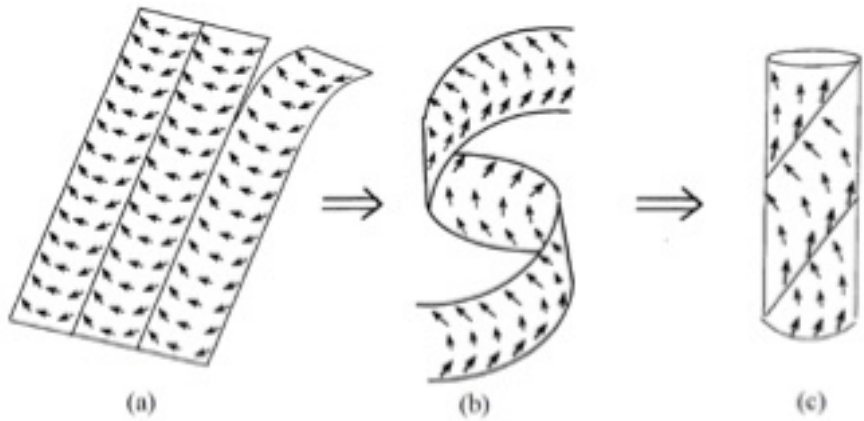

FIG. 12. Schematic illustration of the shape evolution of chiral bilayer membranes (a) with the molecules tilted with respect to their neighbors (indicated by the arrows) twisted into a helical ribbon (b) and a closed tubule when the lipid molecules are deposited from the saturated solution. From Ref. [128], reprinted with permission from AAAS.

### B. Mechanical self-assembly of helical ribbons: microfabrication

Helical ribbons are an important set of geometric shapes among chiral structures. Such shapes often arise as a result of the competition between bending and in-plane stretching energy driven by certain internal or external forces. The sources of driving forces include, but are not limited to, surface stresses [9,135,136], residual stresses [137], misfit strains [10,138-140], molecular tilt [123,129,141], differential growth [17,142], swelling/de-swelling [11,15,23,143] and the coupling between piezoelectricity, electric polarization, and free charge carrier distribution [144].

Strained multilayer structures have received intensive attention from the scientific community due to the applications as sensors and actuators in microelectromechanical/nano-elecromechanical systems (MEMS/NEMS). As mentioned earlier, the mechanical principles of residual stress/strain-induced bending of a multilayer can be exploited to manufacture micro-/nano-scale architectures. Prinz et al. [95] demonstrated that three-dimensional rolled-up nanohelices can be fabricated, through a "top-down" approach, by using an InAs/GaAs bilayer with lattice mis-match strain. The InAs layer is subjected to compression and the GaAs is in tension, because the lattice spacing of InAs is about 7.2% larger than GaAs. The bilayer rolled up (toward GaAs) to partially relax the interlayer strain due to mis-match, after the sacrificial layer (AlAs) was etched. This method relies on the fact that both layers are crystalline materials and grown epitaxially with respect to each other such that the lattice mis-match strain provides the driving force for bending. This is the first report in which the molecular-beam-expitaxy overgrown structures with nanotubes and nanohelices were fabricated. By controlling the geometric orientation of the strip, researchers can control the formation of rings or helices with preferred pitch angles (Fig. 13).

Since then, a variety of different crystalline materials have been used to manufacture bilayer or trilayer nanorings or nanohelices [10,138,145-147]. For instance, Bell et al. [145] manufactured three-dimensional InGaAs/GaAs nanosprings again using AlAs as a sacrificial layer and employing wet etch to release the patterned bilayers. Since crystallographic anisotropy dictates that the <100> direction is the preferred roll-up direction, helical ribbons with different pitches and helix angles formed depending on the misorientation angle of the ribbon's geometric axes and the bending direction (Fig. 13). This fabrication methodology takes advantage of the strain engineering principle discussed above and achieves on-demand manufacturing of semiconductor nanohelices with remarkable precision and reproducibility with potential applications as sensors and microrobotics.

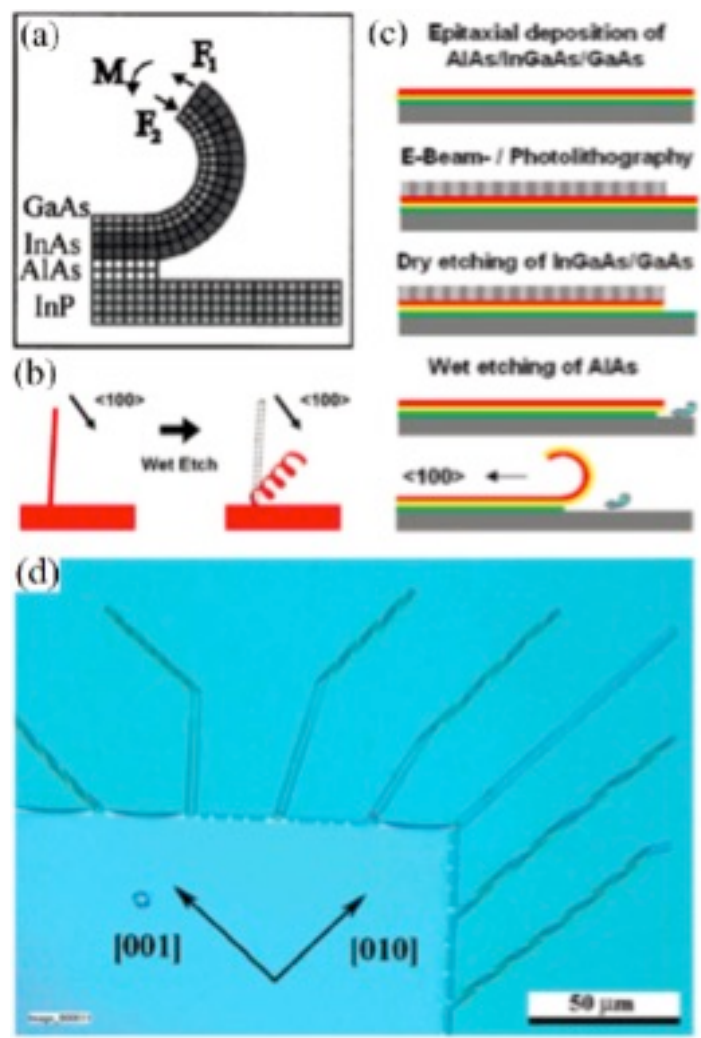

FIG. 13. (a) InGaAs/GaAs strain-induce curl; (b) releasing the strain generates helices; (c) bilayer fabrication; (d) helix pitch is a function of pattern orientation. Reprinted (adapted) with permission from Ref. [145]. Copyright 2006 American Chemical Society.

Zhang et al. [10] further studied the anomalous coiling phenomena in rolled-up SiGe/Si and SiGe/Si/Cr nanohelices. These nanohelices were manufactured using the same methodology as described above. Interestingly, when the width was reduced from 1.3 to 0.7 μm, the pitch and helix angle of the SiGe/Si/Cr helical nanohelix first decreased, then increased, and finally decreased until a self-overlapping ring formed. Noticeably, the chirality also switched from right-handed, to mixed, and to left-handed, suggesting that there could be some edge effects that influenced the morphology. Indeed, Zhang et al. [10] took into consideration the edge effects by hypothesizing that the edge stress would become increasingly dominating when the width decreased, leading to the change of chirality, and the final self-overlapping state when the width fell below a threshold value (Fig. 14). Dai and Shen subsequently used a Cosserat rod theory to interpret this abnormal phenomenon by also considering the increasing edge effects as the width became larger [148].

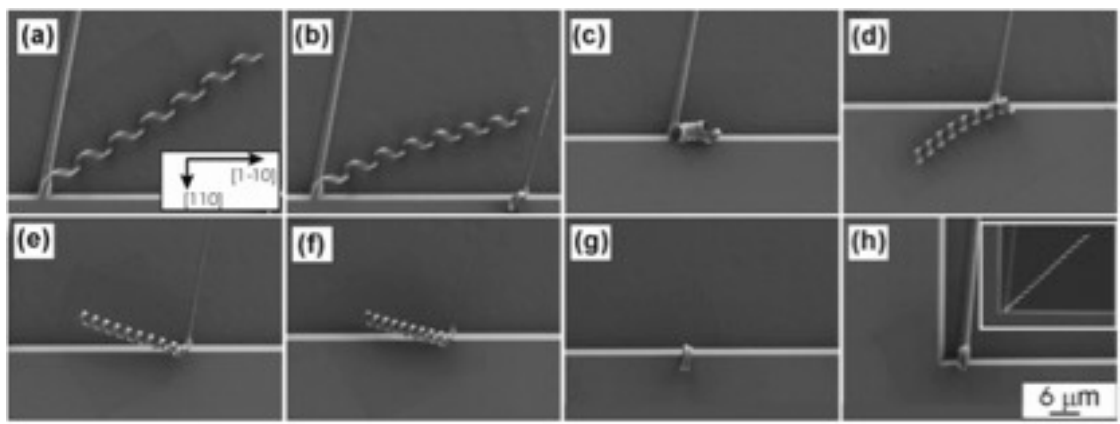

FIG. 14. SEM top view images of SiGe/Si/Cr helical nanoribbons (the layer thickness is 11/8/21 nm). The inset of (h) does not have a Cr layer. The arrows in (a) denote the <110> direction on the substrate. All the strips in (a)-(g) have a misorientation angle of 10° from <110>. The width decreases from 1.30 to 0.70 μm at an interval of 100nm from (a) to (h). In (h), both nanoribbons have a misorientation of 5º. Reprinted (adapted) with permission from Ref. [10] (Copyright 2006 American Chemical Society).

### C. Mechanical self-assembly of helical ribbons: modeling

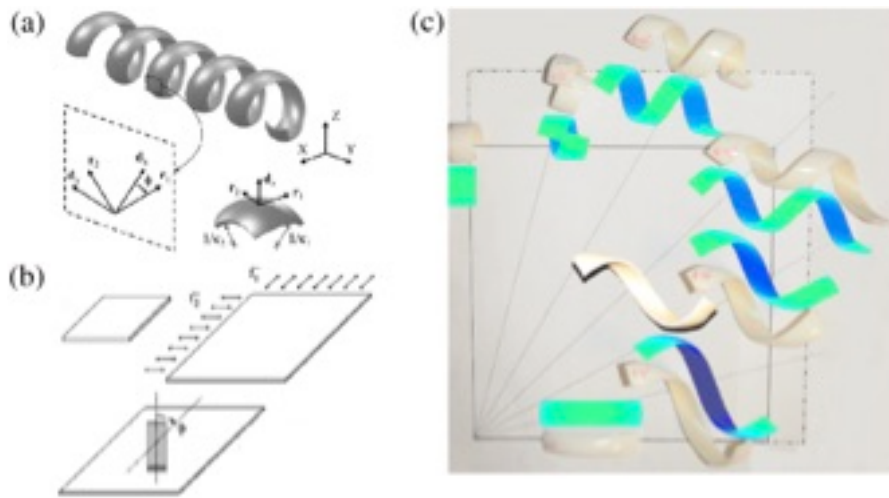

FIG. 15. Tunable helical ribbons. (a) Illustration of a helical ribbon. The principal bending directions, $\mathbf{r}_1$ and $\mathbf{r}_2$, are rotated from the geometric axes, $\mathbf{d}_x$ and $\mathbf{d}_y$, by a misorientation angle $\phi$. (b). (c) a piece of latex rubber sheet is pre-stretched twice as much in the vertical direction than in the horizontal direction before a series of adhesive strips were attached to it along different misorientation angles ($\phi$ =0º, 15º, 30º, 45º, 60º, 75º, and 90º). The ribbon at the center is made of prestretched top and bottom layers with a misorientation angle of 30º. The released multilayer strips deformed into coiled shapes with the pitch and helix angle depending on $\phi$. Reprinted with permission from Ref. [9] (Copyright 2011, AIP Publishing LLC).

More generally, a helical ribbon bends around two principal axes, $\mathbf{r}_1$ and $\mathbf{r}_2$, with principal curvatures, $\kappa_1$ and $\kappa_2$, as shown in Fig. 15a. The principal bending axes form a misorientation angle $\phi$ with the geometric axes $\mathbf{d}_x$ and $\mathbf{d}_y$. Chen et al. [9] showed that the coordinates of the centerline can be parameterized as functions of the arclength s, and the geometric properties, such as the pitch p, the helix angle $\theta$, and the helix radius R, can all be determined by the three independent geometric parameters, $\kappa_1$, $\kappa_2$, and $\phi$:

$$p = 2\pi\tau/\alpha^2, \quad \theta = \sin^{-1}(\tau/\alpha), \quad R = \beta/\alpha^2, \qquad (9)$$

where $\alpha = \sqrt{\kappa_1^2 \cos^2\phi + \kappa_2^2 \sin^2\phi}$, $\beta = \kappa_1 \cos^2\phi + \kappa_2 \sin^2\phi$, and

$$\tau = (\kappa_1 - \kappa_2)\sin\phi\cos\phi. \qquad (10)$$

Further, they considered the deformation of a ribbon driven by surface stress ($f^+$ and $f^-$) acting on the top and bottom surfaces respectively, by minimizing the total energy

$$\Pi = f^- : \gamma|_{z=-H/2} + f^+ : \gamma|_{z=H/2} + \int_{-H/2}^{H/2} \frac{1}{2}\gamma : C : \gamma \, dz, \qquad (11)$$

where $\gamma$ is the strain tensor, C is the elastic stiffness tensor, and the coordinate z is defined as the distance from the mid-plane along the thickness direction. Analytical solutions were then obtained for the case $f^+$ is zero. The theoretical predictions agree well with the table-top experiments (shown in Fig. 15b), in which an elastic strip adhesive is bonded to a pre-stretched latex rubber sheet to produce helical ribbons with different shapes depending on the misorientation angle ranging from 0 to 90° (Fig. 15c).

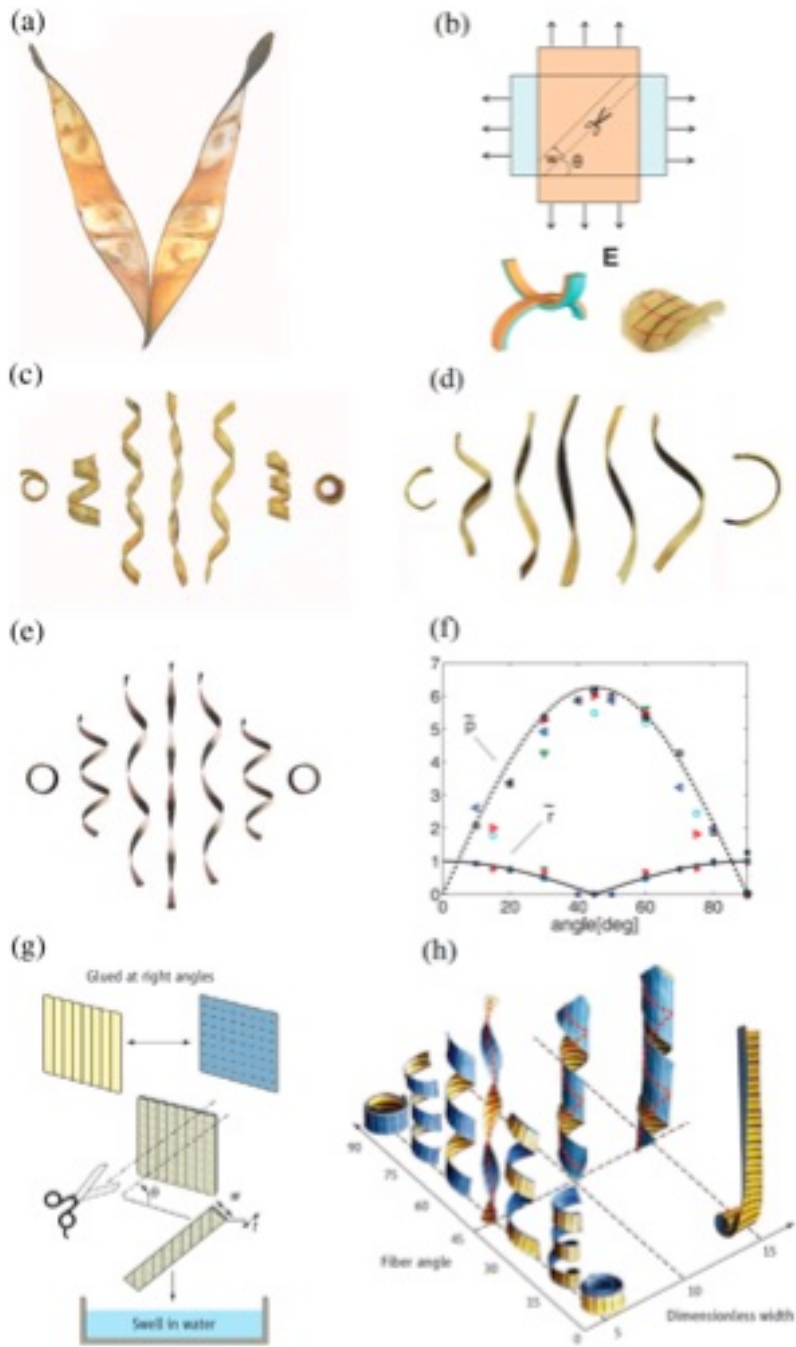

FIG. 16. (a) Open Bauhinia pods. (b) Manufacturing a mechanical analog of Bauhinia pod. Bonding two perpendicularly pre-stretched latex rubber sheets and gluing them together, forms a residually stressed compound sheet. A strip is cut from the sheet along a direction that forms an angle θ with either direction. (c)-(e) Narrow strips cut at angle θ = 0°, 15°, 30°, 45°, 60°, 75° and 90° from a latex sheet, Bauhinia pods and theoretical predictions respectively. (f) Dependence of radius and pitch on angle θ. Symbols correspond

to data points of latex sheet, whereas lines are the theoretical predictions. (a)-(f) are from Ref. [15], reprinted with permission from AAAS. (g) The schematics for the method used to create paper models, which generates the coiling behavior that mimics plant structures. (h) The shapes of paper models for different combinations of the two control parameters: the dimensionless width and the misorientation angle. (g) and (h) are from Ref. [149], reprinted with permission from AAAS.

Armon et al. [15] also developed a theoretical framework to describe the formation of helical ribbons driven by incompatible target metrics and applied the model to interpret the chiral opening of *Bauhinia* seed pods (Fig. 16a). The authors bonded prestretched latex rubber sheets to create a range of helical ribbon shapes. The predicted pitches and radii and the transitions between cylindrical helical ribbons and helicoids (which will be discussed in detail in the next sub-section) agreed well with the resulting ribbon shapes. *Bauhinia* seedpods have been exploiting this mechanism to twist open into two pieces with opposite handedness for millions of years. Forterre and Dumais [149] further used paper models to illustrate the "phase space" of such helical ribbons. The fibers in their papers were typically aligned parallel to one side, so gluing two pieces of papers with perpendicularly aligned fibers formed an anisotropic bilayer. When immersed in water, this bilayer attempted to bend along both directions, creating saddle-shaped structures with negative Gauss curvature. When the strips were cut along different directions, helical ribbons of different shapes formed (Fig. 16h).

Since the theoretical approaches by Armon et al. [15] and Chen et al. [9] do not require the details of any microscopic interactions, the models can predict shape formation from any initially flat elastic sheet driven by residual stresses/strains, surface stresses, swelling/shrinkage or differential growth [149].

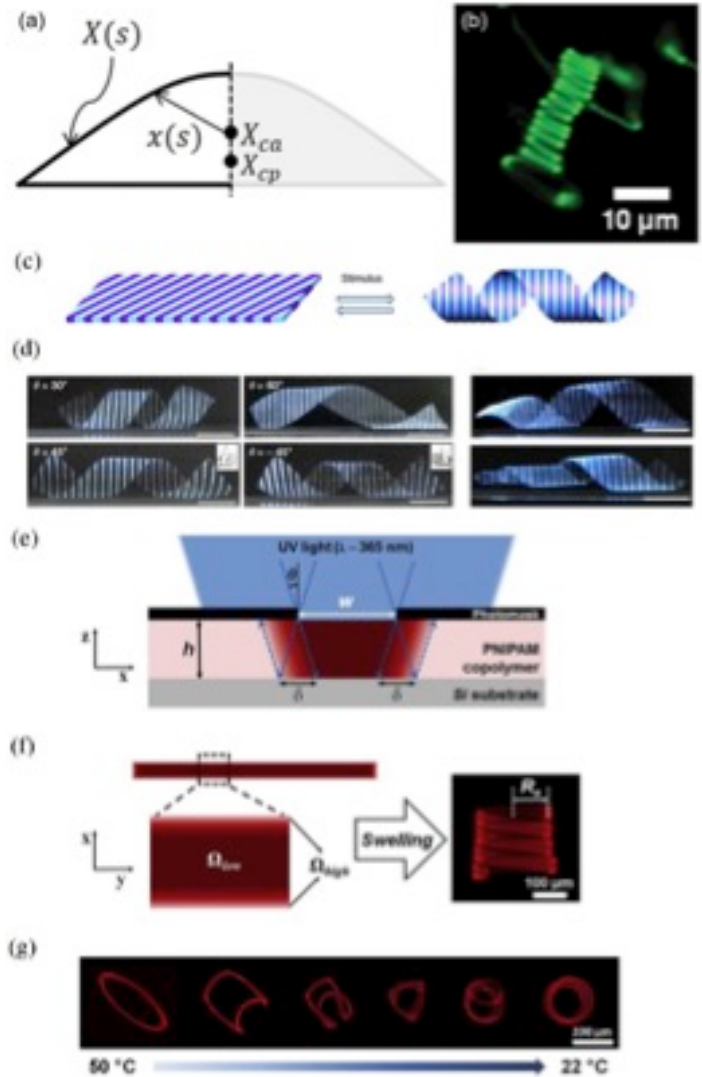

FIG. 17. (a) Theoretical cross-section of isosceles nanoribbons, where $X(s)$ is path of the perimeter. The grey area is the imaginary mirror image. (b) Coiled nanoribbons immersed in water. (a) and (b) are adapted from Ref. [150] with permission. (c) Left panel: a flat sheet made of the composite gel with 1 mm-wide patterned stripes of PNIPAm gel (PG) and PNIPAm/PAMPS gel (BG) that form an angle to the long axis of the sheet. Right panel: A helical ribbon formed when subjected to an external stimulus. (d) Helical ribbons generated after incubating the gel sheets for 24 hours in a 1M NaCl solution. The right panels display conical helices generated in the NaCl solution by gradually changing the geometric parameters such as the ribbon width. (c) and (d) are reprinted by permission form Macmillan Publishers Ltd: Nature Communications, Ref. [151] (Copyright 2013). (e) 365nm UV light is used to selectively crosslink PNIPAM copolymer sheet. The photomask restricts the regions of the sheet that are exposed to light, while the edges of the sheet under the edges of the photomask receive an intensity gradient of light, giving rise to a crosslinking gradient. (f) The swelling constant $\Omega$ is low in the center of the ribbons, indicating that the centers have a limited capacity to swell. The edges are able to swell much more, resulting in a helix with defined radius $R_e$ in a swollen state. (e)-(g) are reprinted from Ref. [152] (Copyright 2014 with permission from Elsevier).

Pham et al. [150] designed and manufactured a class of stretchable nanoribbons able to transform into helices when immersed in water. Ribbons were constructed from metal, polymer, and ceramic materials using evaporative deposition. The nanoribbons have the cross section of an isosceles triangle, and this geometric asymmetry causes the ribbons to spontaneously curl in an attempt to reduce energy (Fig. 17a). The helices are formed only to the degree that they prevent the ribbon from self-intersection, leading to tight coils with low pitch (Fig. 17b). Handedness is governed by the asymmetry of the cross-section.

Helices are common in plant tendrils, and have the tendency to reverse their chirality midway across their length. Godinho *et al.* study this perversion and create temperature-dependent cellulose liquid crystal fibers that mimic this effect [153]. Hydroxypropylcellulose was acetylized and crystalized, then electrospun into active micro and nanofibers. These fibers would wind upon heating while decreasing their tension. The authors also explored how the intrinsic curvature of the electrospun cellulose fibers is a product of fabrication, and that the twist is due to off-core defects [154].

It is also worth mentioning that interfacial energy can also induce helical or even multiple-stranded helical shapes. Ji et al. [155] found that the interfacial adhesion could be responsible for the shaping of some double helices at the microscopic scale. Such double helices have been observed in a variety of systems including DNA [156,157], carbon nanofibers [158], and carbon nanotubes [159,160].

### V. Mechanical principles of shape transitions in self-assembled layers

#### A. Mechanical buckling induces formation of helical ribbons

While mechanical buckling has traditionally been perceived as a failure mechanism, in recent years researchers have employed buckling to construct a variety of geometric shapes including helices. Wu et al. [151] developed a new "small-scale", modulation-based strategy to fabricate two-dimensional sheets that mechanically self-assemble into three-dimensional helical shapes. Inspired by the self-shaping of fibrous organs of plants, they developed stimuli-responsive single-layer composite materials that can undergo shape transformations. More specifically, they patterned a hydrogel sheet with stripes of alternating chemical compositions at a misorientation angle to the geometric axis of the ribbon. The difference in the swelling/de-swelling ratios and elastic moduli between alternating stripes leads to a shape transformation of the 2D sheet into a helical ribbon, driven by a reduction in stretching energy. Remarkably, instead of having a residual strain/stress gradient (often through a multilayer design), as was the case in previous works, the single layer in this study meant that there was initially zero elastic modulus and stress/strain gradient along the thickness direction. In fact, the formation of helical ribbons here is due to mechanical buckling associated with the release of in-plane stretching energy, so the bending direction is always along the direction of the stripes. Cylindrical helical ribbons with both right-handedness and left-handedness were generated with equal probability (Fig. 17c).

The buckling of thin hydrogel sheets has also been shown to be controllable and reversible, paving the way for designing materials that respond naturally to their environment. Bae et al. [152] developed a simple manufacturing method to induce helical self-assembly when exposed to changes in temperature. Rather than changing the chemical composition of hydrogel sheets at different points, the researchers photo-crosslinked the gel in such a way as to limit crosslinking on the edges of the ribbon (Fig. 17e-f). The edges of the resulting ribbon were thus able to swell more than the center. Like the hydrogel ribbon developed by Wu et al. [151], the ribbon releases in-plane stretching energy by curling around a radius $R_e$ (governed by the width of the ribbons). The curling of these ribbons was reliably reversible when the ambient temperature was increased to 50°. At these temperatures, the ribbons were less able to swell and the resulting in-plane stress between the edge and center was lower, reducing the curvature of bending. The authors posited that this behavior could have applications as a micro-knot that can be tightened or loosened in response to temperature. It should be noted that, unlike the periodically patterned hydrogel sheets in Wu et al. [151], a hydrogel ribbon that curls under edge effects alone has a natural pitch of 0 – i.e. it has no off-axis bias will naturally form a helix only to the extent that it prevents self-intersection (Fig. 17b). A bias could theoretically be introduced using the off-axis stripe techniques of Wu et al. [151], nevertheless. Interestingly, the authors found that the side of the ribbon exposed directly to UV light during photo-crosslinking would reliably end up as the inside of the curled ribbon. This was because the up/down orientation of the ribbon with respect to the light introduced a crosslinking gradient in the z-direction. The side closer to the light cross-linked more, making it less able to swell and pushing it to the inside of the curl.

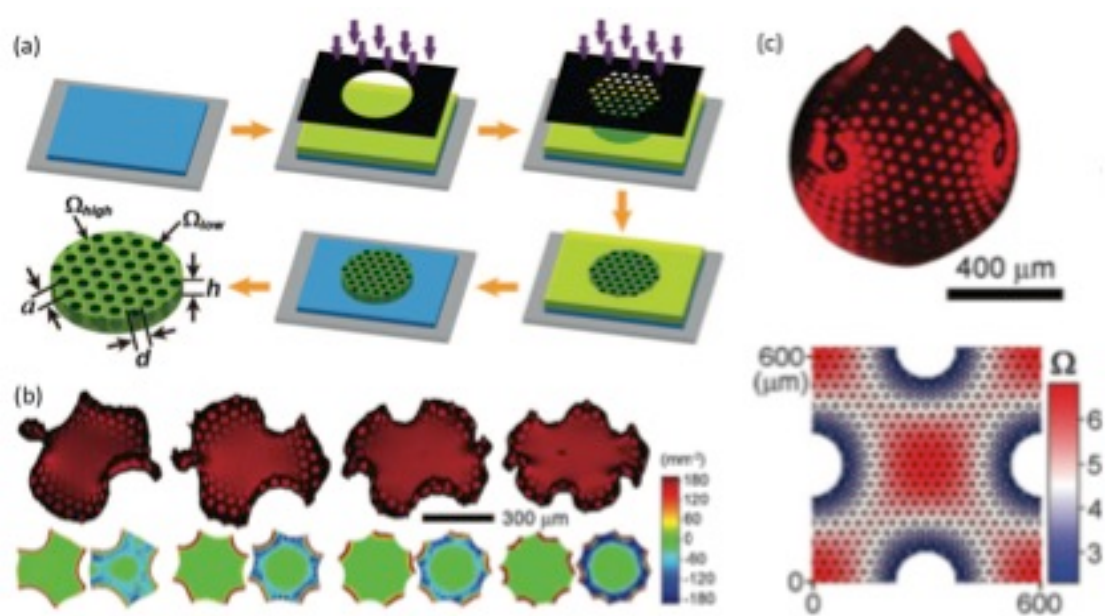

FIG. 18. (a) Flowchart showing the halftone lithographic patterning method. A simple

photomask is used to give the entire region a low dose of UV, and then a halftone photomask gives selective high doses of UV. (b) Enneper's minimal surfaces generated from radially symmetric patterning. Surfaces with 3 to 6 nodes are generated. (c) Sphere generated from asymmetric patterning, with areas of high distortion excised. From Ref. [161], reprinted with permission from AAAS.

While crosslinking-gradient edge effects serve as an extremely simple way to manufacture shape-morphing sheets, significantly more complicated structures can be built by selectively crosslinking the whole surface of a sheet. Kim et al. [161] used a "halftone lithography" approach to achieve arbitrary swelling and shrinking of sheets in two dimensions. They created an acrylic acid monomer solution with benzophenone photo-crosslinking units that gels proportionally to the amount of UV light received, allowing smooth variations in crosslinking levels. Rather than use difficult-to-produce grayscale photomasks, the authors opted to simulate gradations through the use of a halftone process, where the density of tiny circular regions of high UV is used to simulate varying levels of exposure (Fig. 18a). This gives a wide range of possible crosslinking levels with only two binary photomasks. With this method, the authors first showed that it was possible to easily produce common shapes that require only radially symmetric patterning such as saddles, cones, and Enneper's minimal surfaces (Fig. 18b). These minimal surfaces were formed by mapping the swelling equation:

$$\Omega(r) = c\left[1 + \left(\frac{r}{R}\right)^{2(n-1)}\right]^2, \tag{12}$$

where n is the number of nodes and $\Omega$ is the swelling ratio. It was found that the number of nodes on the resulting 3D shape could indeed be predictable varied, showing good agreement with theory. More interesting shapes are possible with patterning that varies in two dimensions, such as a sphere. The authors took a conformal mapping of a square onto a sphere:

$$\Omega(x,y) = 2\frac{|dn(\alpha)sn(\alpha)|^2}{[1+|cn(\alpha)|^2]^2}, \alpha = \frac{x+iy}{R}|\frac{1}{\sqrt{2}} \text{ where } dn, sn, cn \text{ are Jacobi elleptic functions}$$

This mapping has the advantage of concentrating distortion (which corresponds to extremely high or low levels of crosslinking) to a few small points, which could simply be excised from the sheet before swelling. The resulting (Fig. 18c) approximates a sphere, though the corners do not meet because of singularity effects.

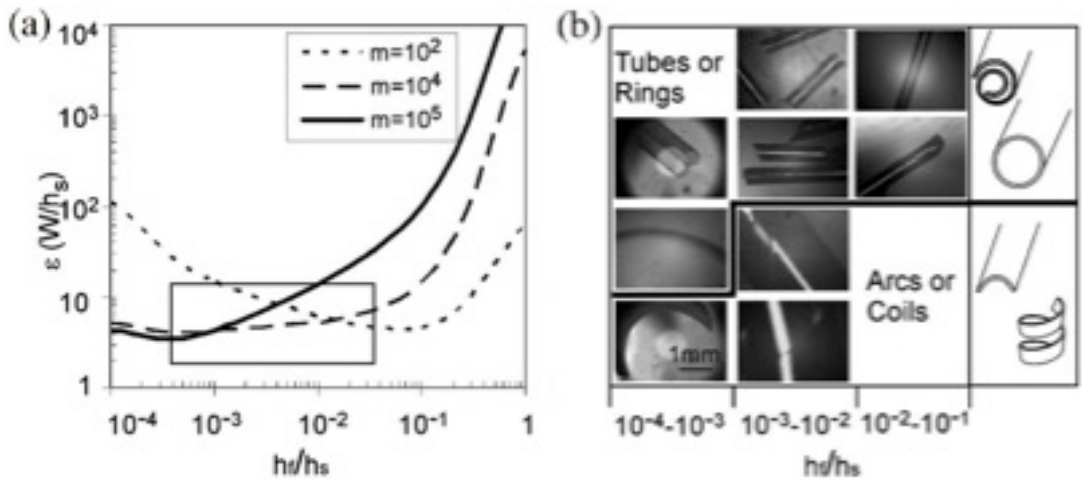

FIG. 19. (a) Phase diagram showing the cutoff point for when polymer bilayer particles will form complete tubes and rings (above lines) versus arcs and coils (below lines). *x*-axis is ratio of layer thicknesses, *y*-axis is radius of curvature, and each line corresponds to a different biaxial modulus. (b) Examples of structure found above and below the lines. (a) and (b) are reprinted with permission from [162] (Copyright 2008, AIP Publishing LLC).

There has been significant research into the development of PDMS based adaptive bilayers that respond naturally to changes in environmental conditions, since PDMS's high biocompatibility makes it a promising candidate for drug delivery. Simpson et al. [163] shows that when plated with a dissimilar material, such as gold, the interfacial tension between the two materials causes controlled wrinkling. This wrinkling can be directed by locally modifying the thickness of either layer, leading to the creation of complex structures. The bilayers that they created were able to coil and uncoil in response to temperature changes, allowing them to capture and release poly(ethylene glycol). In a similar study, Kalaitzidou and Crosby [162] showed that it was possible to create adaptive polymer bilayer particles that underwent shape actuation when exposed to interfacial stress. These small particles changed the characteristics of the fluid flows they were in based on whether they were in a rolled or unrolled state. Sheets of PDMS were cured and plated with a gold layer through e-beam evaporation. These sheets were then cut into pieces with set lengths and widths, which determined their subsequent shape transformation. They found that when both dimensions are significantly larger than the PDMS sheet thickness, the particles tended to roll into tubes and rings. As the thickness increased, the particles rolled into open structures such as arcs and helices (Fig. 19a-b). This process was shown to be reversible, with the coils and tubes returning to a two-dimensional state upon sufficient temperature increase. They showed that the small size of the coils allow them to be used as adaptive particles in solution to control the flow of that solution based on temperature. This is generalizable to a large close of responsive polymers and environmental stimuli.

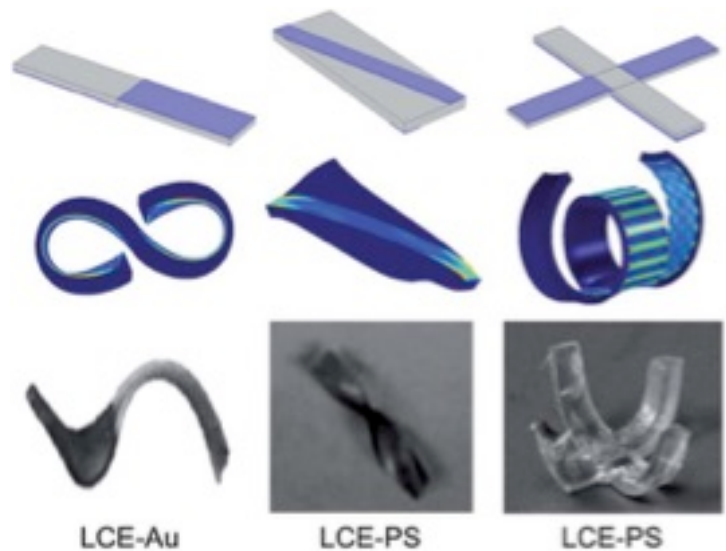

FIG. 20. Various structures formed by LC elastomers adhered to a polystyrene layer. Bends, folds, and twists were introduced, as well as a four-arm grabbing structure. Reproduced (adapted) from Ref. [164] with permission of The Royal Society of Chemistry.

Agrawal et al. [164] showed that uniformly aligned liquid crystal elastomers could be made to actuate in complex shapes by affixing a secondary polystyrene layer that locally varies in thickness. When heated, the LC elastomer attempts to contract along its alignment director but is inhibited by the polystyrene layer. In regions of thin polystyrene the balancing of deformation energy causes small wrinkles, but as the polystyrene becomes thicker, the wavelength and amplitude of the wrinkles increase until the sheet folds. The authors show that applying polystyrene films on opposite layers can make further complex shapes. A four-arm grabbing actuator was made by selectively patterning the arms of a cross (Fig. 20). The polystyrene is placed on top of the LC where the alignment director is parallel to the direction of the intended bend, while it is place under the LC with perpendicular alignment. When heated, both sets of arms curl upward in a grasping motion. This pattern can be extended to a planar LC elastomer / polystyrene bilayer, which will result in an actuator that functions like a leaf – closing when the temperature rises too high and opening when it falls.

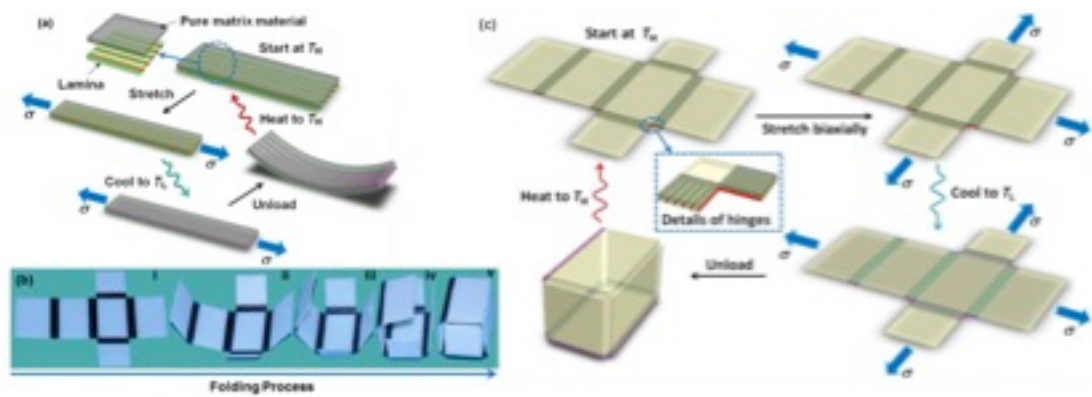

FIG. 21. a) Thermal cycle of shape memory fiber-embedded elastomers. When stretched and cooled below glass transition temperature of fibers, the composite bends on unloading. When reheated, it returns to a flat state. b-c) Self-folding box made from this technique, where the fibers are embedded in the folds. Reprinted with permission from Ref. [5] (Copyright 2013, AIP Publishing LLC).

Shape memory alloys have long been a promising material in the field of shape-morphable structures but have been hindered by their lack of flexibility. Recent advances are helping to remove this barrier, such as the work of Ge et al. [165], who designed a composite structure where shape memory fibers were embedded in an elastomer matrix [165]. This allowed them to encode shape-morphing behavior into the fibers. Using a multimaterial 3D printer, glassy shape-memory fibers with preprogrammed shape effects were printed into the elastic matrix. When the resulting material is stretched and cooled below the glass transition temperature of the fibers, it curls when the stretching force is released. When the heated to above the glass transition temperature, the material recovers its original shape (Fig. 21a). The authors used this property to design and print a self-folding box, in which each of the hinges were composed of this curling design. The box is stretched and cooled and upon release folds itself. When reheated, it similarly unfolds (Fig. 21b-c).

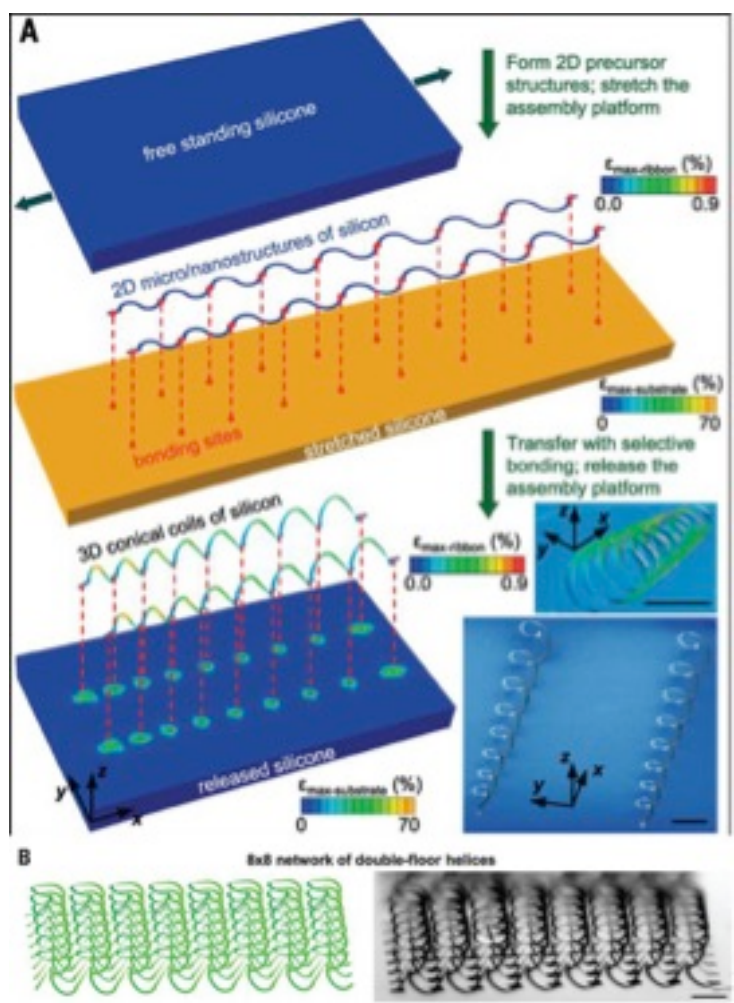

FIG. 22. a) Nano-helix structure formed from wavy silicon ribbon selectively bonded to elastomer sheet. b) A similar technique is used to make double helix structures, where the second layer of helices stands on top of the first. This second layer is not bonded to the elastomer. From Ref. [166], reprinted with permission from AAAS.

Xu et al. [166] have shown that it is possible to induce buckling-driven self-assembly of helical ribbons in almost arbitrary materials. Winding silicon ribbons were placed on a prestretched elastomer. Using UV light to generate ozone at specific points on the surface of the ribbon induces the formation of hydroxyl groups, which allows the ribbon to bind to the prestretched elastomer at well-defined points. When the elastomer is allowed to relax, the ribbons deform out of plane in order to minimize internal stress (Fig. 22a). In this case the radius of the resulting helix, as well as the pitch and chirality, is all determined by the spacing of the bonding hydroxyl groups in relation to the ribbon. Unlike the helices made of hydrogels in Bae et al.'s work [152], these helices once formed are permanent. The authors also found that by using finite element analysis to pre-compute the strain on the ribbons they were able to manufacture multilayer architectures. The higher layer strips of silicon, rather than buckling around hydroxyl bonding sites, would buckle around their joins to other silicon strips, enabling the formation of double-layered helices (Fig. 22b). This new technique is very promising primarily because of the impressive maturity of silicon lithography technology. Since the deformations rely on the same techniques that are used to pattern silicon chips, 3D out-of-plane nanostructures can be made at extremely high resolution. The authors showed that this technique was also extensible to metals and other semiconductors, indicating impressive versatility.

## B. Shape transitions in helical ribbons

Recent studies have shown that shape transitions can occur between purely twisted ribbons, helicoids, spiral ribbons, and tubules by changing the relevant geometric parameters, such as the magnitudes and signs of the principal curvatures, the misorientation angle, or the geometric dimensions [9,15,21-23]. These changes often result from external stimuli, such as changes in temperature[24,167], pH, or swelling/deswelling [15,23].

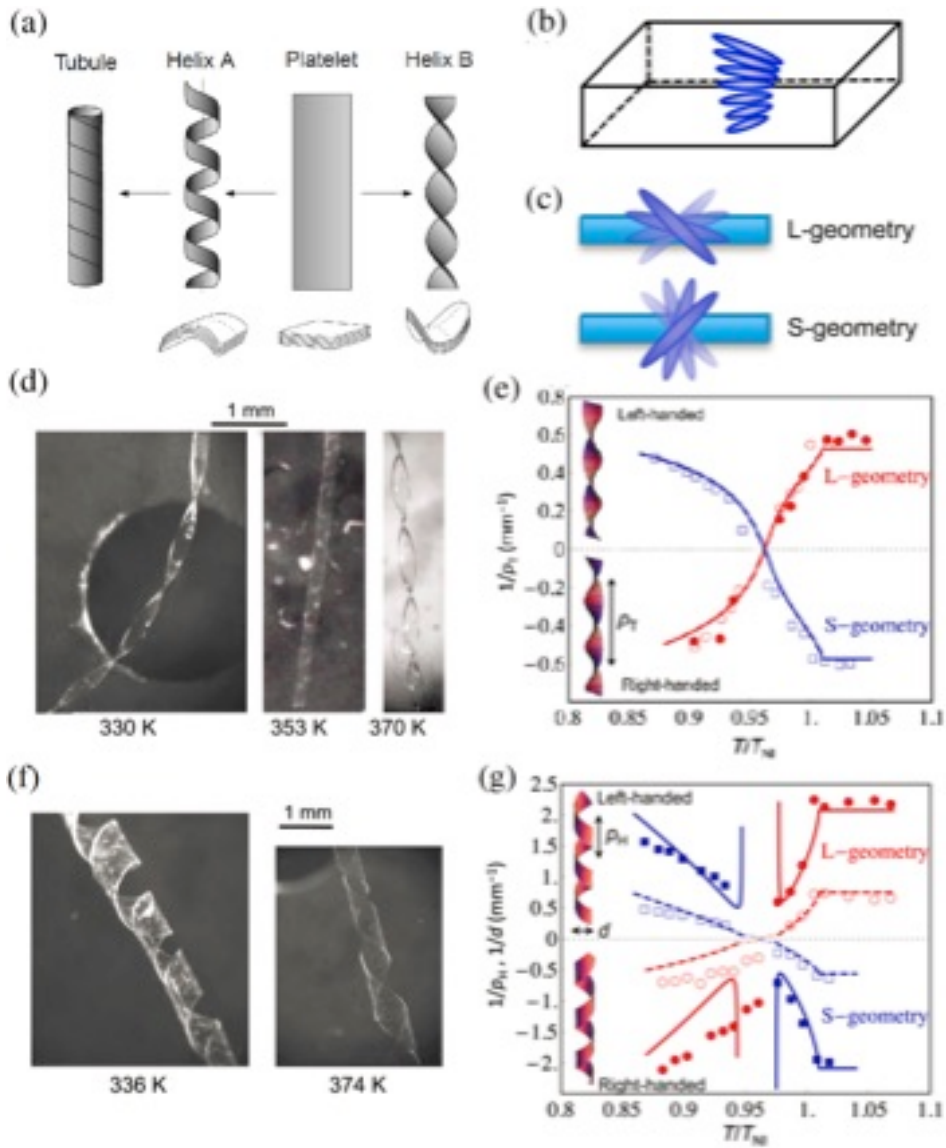

FIG. 23. (a) Schematic representation of cylindrical helical and purely twisted ribbons. The bottom shapes feature the locally cylindrical and saddle-like curvatures in these multilayered ribbons. Reprinted by permission from Macmillan Publishers Ltd: Nature, Ref. [127] (Copyright 1999). (b) Side view of schematics of the direction of TNE ribbons. (c) A schematic diagram shows the top view of L- and S- geometry. (e) Temperature dependence of the inverses of pitch and diameter of the helices. The lines (both dashed and solid) represent theoretical predictions. The (red) circles and (blue) squares represent the data of the L-geometry and the data of the S-geometry respectively. (f) Cylindrical helical ribbons formed

by the wide twisted-nematic-elastomer (TNE) where the thickness of the ribbons is 35.2 $\mu m$. The L-geometry ribbon is left handed at 374 K and right handed at 336 K. Adapted from Ref. [24] with permission.

Oda et al. [127] studied the transition between helicoids and spiral ribbons in charged gemini surfactants when the length of the molecular chain changed (Fig. 23a). These shape transitions were also observed in twist-nematic-elastomers [1,24]. Sawa et al. [24] found that thin ribbons underwent a similar transition when the width varied and developed a theoretical model by introducing a term in the total potential energy that accounts for molecular twist. Twist-nematic-elastomer ribbons can form purely twisted or cylindrical helical shapes depending on the temperature and/or width. This shape selection arises from the competition between bending energy and in-plane stretching energy. A ribbon with a small width-to-thickness ratio can easily adopt a twisted shape with a large Gauss curvature; while one with a large width-to-thickness ratio will stay in a shape with nearly zero Gauss curvature to minimize the total potential energy (and in particular the stretching energy). The change in temperature, on the other hand, results in local stretches (or shrinks) along the nematic director caused by residual effects from temporary chiral dopants. They also showed that certain liquid crystal elastomer ribbons could switch chirality when transitioning with changes in temperature.

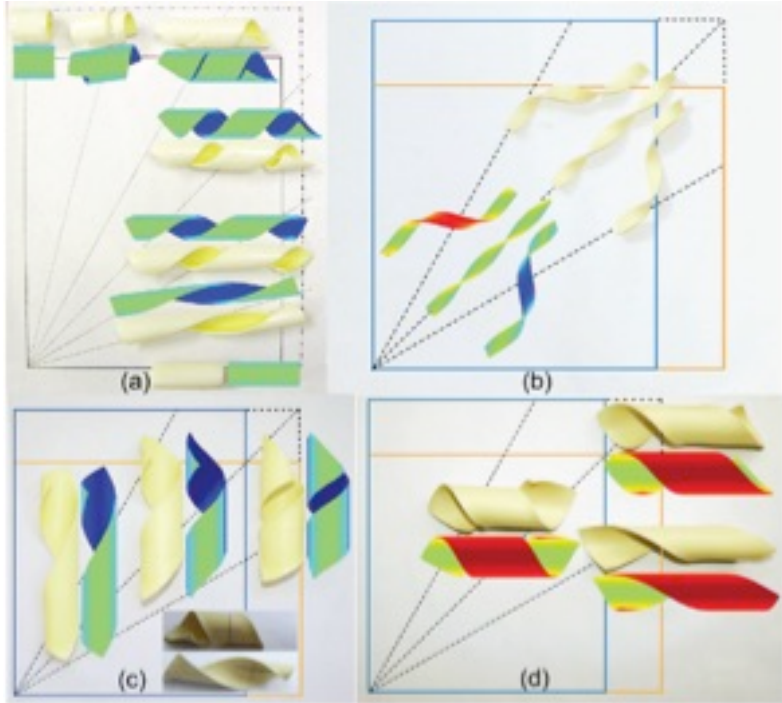

FIG. 24. Monostable and bistable helices. (a) Mono-stable helical ribbons. An elastic strip was bonded to a prestretched latex rubber sheet, with a misorientation angle $\phi$ ranging from 0 to 90º at a 30º interval. The pre-stretches are 0.24 and 0.12 in the vertical and horizontal directions, respectively. (b) Mono-stable helical ribbons. (c) Bistable-helical ribbons. (d) The other stable shapes of the same ribbons in (c). The color is indexed according to the out-of-plane displacement. Reprinted with permission from Ref. [22] (Copyright 2014, AIP Publishing LLC).

Armon et al. [15,23] and Guo et al. [22] independently examined shape transitions in strain-engineered elastic ribbons and came up with similar criteria that a transition would occur at when the dimensionless "width" $W/\sqrt{\kappa/H}$ (where $W$ is the width, $H$ is the thickness, and $\kappa$ is the principal curvature) exceeds the threshold value. So the transition in shape (as well as multistability, as will be discussed later) is actually dictated by the combination of these three geometric parameters. Guo et al. [22] further performed theoretical analysis and experiments (Fig. 24) to study such shape transitions and the associated change in multistability. Finite-element simulations (Fig. 25) were also employed to quantitatively investigate such shape transitions [168]. (In fact, these shape transitions not only occur in surfactants, strained elastic ribbons, and liquid crystal elastomers, but also in seedpods and peptides related to Alzheimer's diseases.)

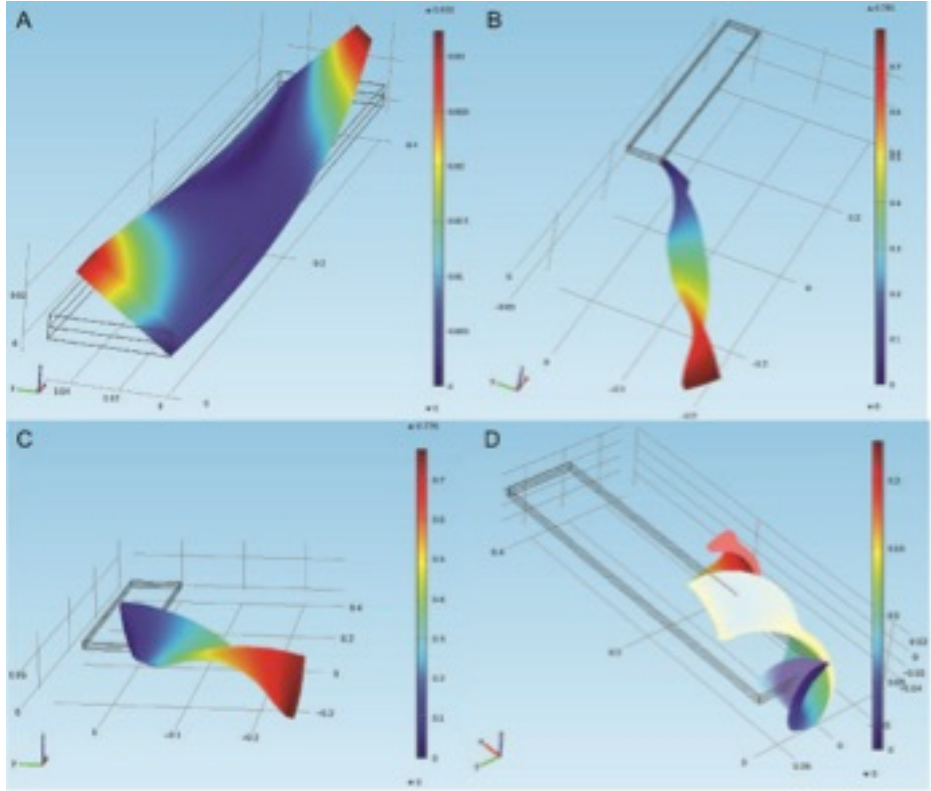

FIG. 25. The stable helical configurations driven by misfit strains in bilayer ribbons modeled using finite element simulations. The misfit strain in the top layer is

$$\underline{\underline{\varepsilon}}_t^0 = (\varepsilon_1 \cos^2\phi + \varepsilon_2 \sin^2\phi)\underline{e}_1 \otimes \underline{e}_1 + (\varepsilon_1 \sin^2\phi + \varepsilon_2 \cos^2\phi)\underline{e}_2 \otimes \underline{e}_2 + (\varepsilon_1 - \varepsilon_2)\sin\phi\cos\phi(\underline{e}_1 \otimes \underline{e}_2 + \underline{e}_2 \otimes \underline{e}_1)$$

, where (A) $\varepsilon_1 = 0.02$, $\varepsilon_2 = -0.02$, $\phi = \pi/4$, $L = 0.4\,\text{m}$, and $W = 0.04\,\text{m}$; (B) $\varepsilon_1 = 0.1$,

$\varepsilon_2 = -0.1$, $\phi = \pi/4$, and $W = 0.04\,\text{m}$; (C) $\varepsilon_1 = 0.1$, $\varepsilon_2 = -0.1$, $\phi = \pi/4$, $L = 0.4\,\text{m}$, and $W = 0.08\,\text{m}$; (D) $\varepsilon_1 = 0.5$, $\varepsilon_2 = -0.5$, $\phi = \pi/4$, $L = 0.4\,\text{m}$, and $W = 0.08\,\text{m}$. Transition from a purely twisted ribbon (C) to a spiral ribbon (D) occurs when the misfit strain is increased by five times. The color (in this figure and the following) is indexed according to the total displacement to help visualize the deformation involved. With kind permission from Springer Science+Business Media, Ref. [168] (Copyright 2014).

Among the key parameters that control the shape of the deformed ribbon are the misfit strains, the misorientation angle, the width, the thickness, and the elastic layer properties. Notably, a purely twisted ribbon forms (Fig. 24b and Fig. 25b) when the principal misfit strains are such that

$$\varepsilon_1 \cos^2 \phi + \varepsilon_2 \sin^2 \phi = 0. \tag{13}$$

This result agrees well with the previous study by Chen et al. [9], namely, a purely twisted ribbon forms if and only if $\kappa_1 \cos^2 \phi + \kappa_2 \sin^2 \phi = 0$ (where $\kappa_1$ and $\kappa_2$ are the principal curvatures).

It is worth pointing out the differences between a helicoid and a purely twisted ribbon, which can be confusing at times [169]. A helicoid is the only ruled minimal surface besides a plane. In fact, the centerline of a helicoid ribbon does not necessarily have to be straight, but can also be a helix. In comparison, a purely twisted ribbon has a straight centerline and the radius of the bounding cylinder is zero. Mathematically, a helicoid has principle curvatures $\kappa_1$ and $\kappa_2$ such that $\kappa_1 + \kappa_2 = 0$. Therefore, a purely twisted ribbon becomes a helicoid when $\kappa_1 = -\kappa_2$ and $\phi = \pi/4$.

Cranford and Buehler [170] developed a multiscale molecular dynamics approach by adopting a two-dimensional coarse-grained model to recapitulate the mechanical self-assembly of mono- and multi- layer graphenes. A spontaneous transition from a purely twisted configuration (helicoid) to a coiled shape in graphene ribbons was identified, which gave rise to a strain filled with more homogeneity. The results are similar (but not identical) to the experimental observations and theoretical predictions by Armon et al. [15,23] and Guo et al. [22], as well as the results by Lee et al. [171], Kit et al. [172], and Chen et al. [137], but in the work by Cranford and Buehler, the transition to the coiled shape is mainly because of a mechanical instability "between the imposed strain of the twisted graphene ribbons and the bending stiffness" [170]. As a result, this transition occurs more frequently in stiffer or thicker graphene ribbons.

Recently, Wu et al. [151] demonstrated the possibility of achieving multiple "programmed" shape transformations between different geometric shapes under ambient conditions by integrating multiple, small-scale modulated structural components within planar sheets. Continuous, reversible shape transformations in response to external thermal or chemical stimuli, from planar sheets to three-dimensional arcs, to helices, and tubules, were fulfilled by this novel design strategy which exploits mechanical buckling principles. This approach can be readily extended to other materials such as elastomers and liquid crystalline polymers to enable new functionality that can be coupled with shape transitions.

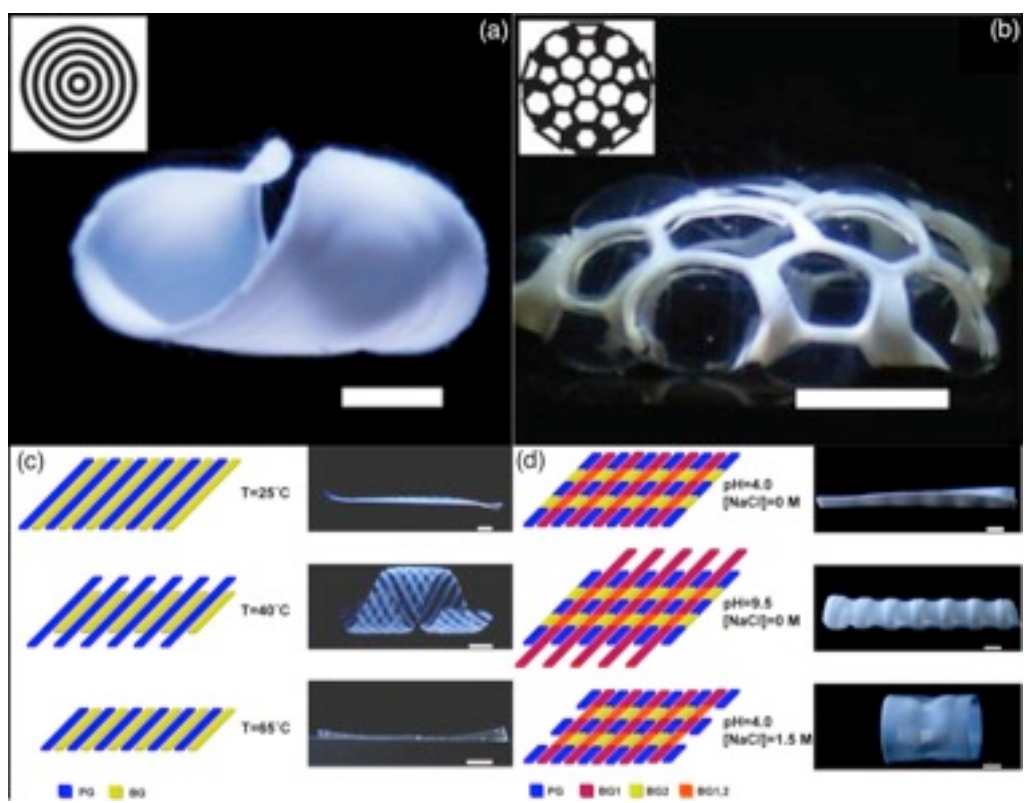

FIG. 26. (a) Saddle-shape structure formed from concentric ring-patterned sheet. b) Icosahedron formed from projection-patterned sheet. c) Multi-stage material that coils upon reaching 40° and uncoils upon reaching 65°. d) Multi-stimulus material that curls into a long thin tube under high pH and a short fat tube under high salinity. Reprinted (adapted) with permission from Ref. [173] (Copyright 2013 American Chemical Society).

Therien-Aubin et al. [173] developed a hydrogel able go through multiple shape transformations in response to a wide range of stimuli, including temperature, pH, ionic strength, and $CO_2$. These gels also responded differently to various stimuli. They first showed that a sheet treated with poly(N-isopropylamide) and patterned with a truncated icosahedron would assume the shape of an icosahedron when exposed to a NaCl solution. By using a different chemical treatment, they were able to force a sheet printed with circles of chemicals to adopt a saddle shape when exposed to $CO_2$ (Fig. 26a), and an icosahedron-patterned sheet to adopt a similar structure (Fig. 26b). These transformations were reversible and showed resilience to hysteresis. The authors then patterned a hydrogel ribbon to respond to heat: as it was heated above ambient temperature, it curled into a helix, but when heated above the lower critical solution temperature of the doping chemical, it returned to a flattened state (Fig. 26c). Finally, a hydrogel sheet was manufactured that responded to both pH and ionic concentration. The sheet was patterned with stripes along the width that activated under changes in pH, while stripes along the length moved in response to changes in ambient ionic concentration. This resulted in a gel that was able to assume a gradient of shapes based on the relative strengths of the two stimuli: when exposed to high pH alone, the sheet rolled into a tight long cylinder, and when exposed to high ionic concentrations alone, the sheet rolled into a wide compressed cylinder (Fig. 26d).

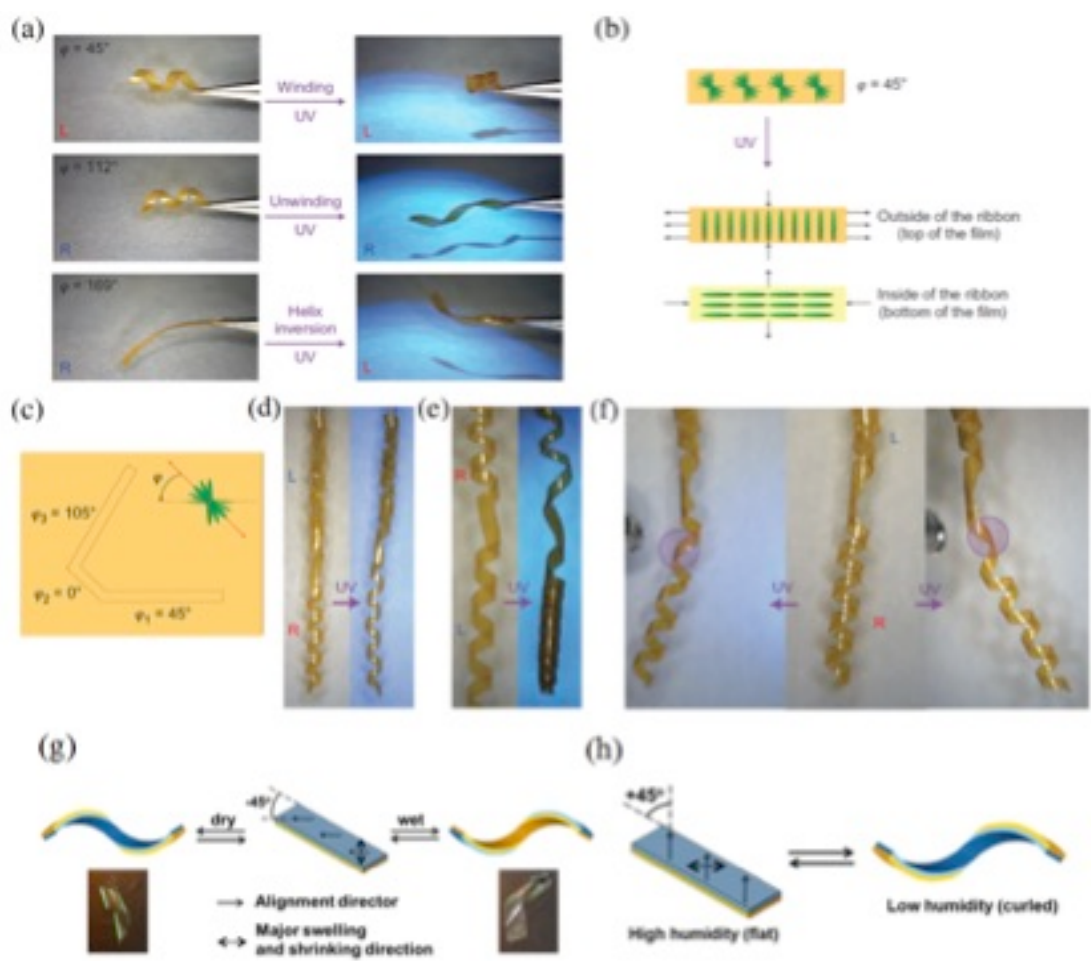

FIG. 27. (a-b) LC helix curling, uncurling and inversion under UV light exposure. Helices where the handedness is reinforced by the alignment director will curl when exposed to UV light, but those where handedness is opposite to alignment director will uncurl and eventually invert. (c-f) When selectively irradiated by UV light near the kink, the end of the helical ribbon constructed from liquid crystal polymer moves away and off axis from the other end. Reprinted by permission from Macmillan Publishers Ltd: Nature Chemistry, Ref. [174] (Copyright 2014). (g-h) Diagram showing how alignment director of the initial LC network informs the handedness of the resulting helix. (g) and (h) are reprinted (adapted) with permission from Ref. [175] (Copyright 2014 American Chemical Society).

Furthermore, the chemical-mechanical coupling in these model systems may advance the current understanding of the deformation and actuation of fibrous living organs [151]. In particular, Iamsaard et al. [174] showed that helical ribbons constructed from liquid crystal polymer networks could be designed to mimic the response of plant tendrils unfolding (Fig. 27a-f). Current research indicated that the elongation of rectangular plant cells in one direction and shrinking of the cells in the transverse directions to cause tendril unwinding is very similar to the shape-change of liquid crystals under stimulation [153]. The authors attached a helical coil with right-hand chirality to a coil with left-hand chirality, joined at a kink (the equivalent to a tendril perversion in plants). When selectively irradiated near the kink by UV light, side belonging to the right-handed helix coiled further while the side belonging to the left-handed helix uncoiled (Fig. 27f). This behavior caused the end of the synthetic tendril to move both away and off axis from the other end, as the kink became a joint where macroscopic bending can take place.

Liquid crystal polymer networks have also been made to actuate in response to changes in humidity. De Haan et al. [175] constructed a LC polymer sheet that could undergo bending and twisting when stimulated by uniform humidity (Fig. 27g-h). After dipping in KOH and rinsing in water, the sheet bent towards the untreated side, but when dried it bent towards the other side (Fig. 27f).  The authors treated the sheets with basic solution in specific patterns and observed the resulting shape change in humidity. A ribbon treated with an alternating pattern assumed an accordion shape in low humidity, while a ribbon with narrow treated bands formed a sharp hinge. They then cut a ribbon with an alignment 45° off-center and showed that it curled right-handed around an axis in the dry state and became straight in high humidity, forming helices (Fig. 27g).

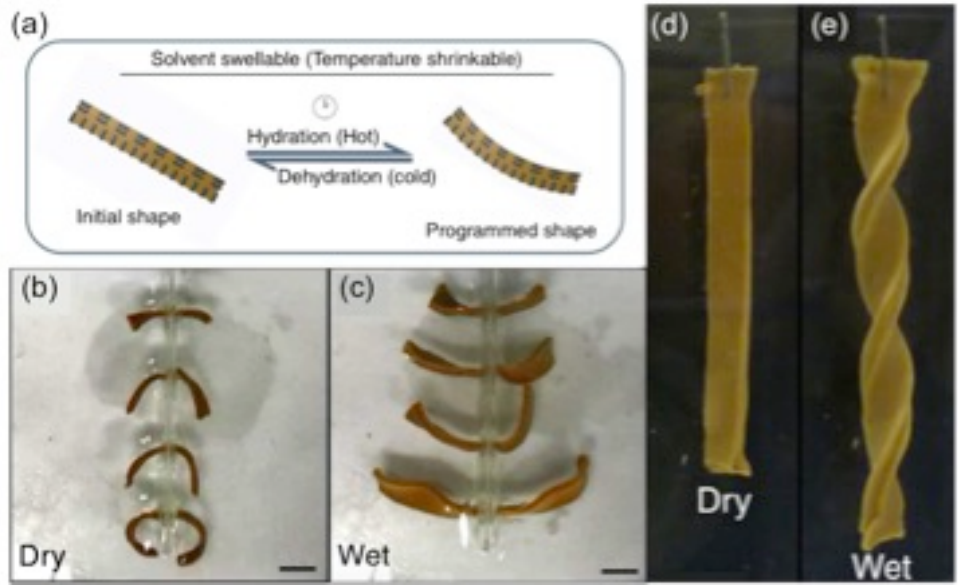

FIG. 28. (a) Diagram showing programmed shape transition under hydration. (b), (c) Dry and wet states of uniform curling and uncurling composites. (d), (e) Dry and wet states of twisting composites. Reprinted by permission from Macmillan Publishers Ltd: Nature Communications, Ref. [176] (Copyright 2013).

Plant shape change, on the other hand, is generally driven by uniform changes in environmental conditions, such as humidity. Thus, any asymmetric shape-change must be encoded in the plant's internal heterogeneous structure, usually through the judicious placement of stiff cellulose microfibrils (CMFs) that respond differently to humidity than the surrounding tissue. Erb et al. [176] developed a framework for manufacturing synthetic shape-change composites based on this natural design. The authors investigated the reinforcement architecture seen in seed dispersal units that bend and twist. Aluminum oxide platelets electrostatically bonded to super-paramagnetic iron oxide nanoparticles were used to simulate the effect of CMFs while also allowing the orientation of the platelets to be controlled by weak external magnetic fields. These platelets were mixed into fluid polymer solutions to produce bulk hygroscopic composites, and the resulting solutions were gelled in the presence of magnetic fields. Composites that bend were produced using a bilayer configuration to mimic pinecone architecture, in which the first layer was deposited under the influence of a uniform magnetic field, and the second from a rotated uniform field (Fig. 28a). When the resulting composite was dried, it bent towards the layer with the platelets oriented perpendicular to the long axis (Fig. 28b-c). A similar system that twisted like a seed dispersal unit was built by changing the bending bilayer configuration to have all the platelets also off-axis by 45°. This caused the layers to attempt to expand in directions perpendicular to one another and results in a twist (Fig. 28d-e). Both the bending and twisting systems could be brought back to a flat state through drying. Multi-responsive composites were also obtained that responded to both hydration and heat. Hydrogel with alumina platelets in the twisting bilayer configuration would twist to the right when heated as the gel lost its ability to store water. When hydrated over a long period of time, the composite would twist with the opposite chirality.

Many of the physical systems mentioned above have been inspired by biological phenomena. In turn, systematic studies on the chemical-mechanical coupling in these model systems may advance the current understanding of the deformation and actuation of fibrous living organs [151].

## C. Change of handedness

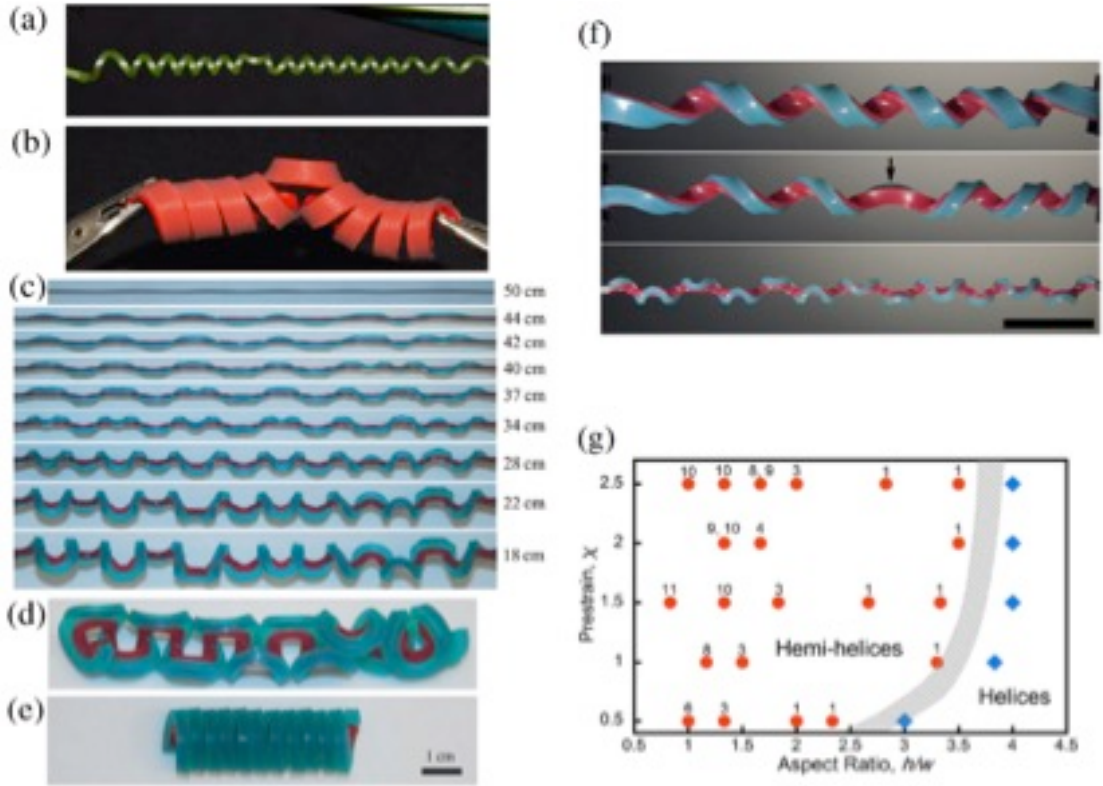

FIG. 29. (a) Fiber ribbon that overwinds around the perversion when pulled. (b) Twistless spring with low bending stiffness and high twisting stiffness that unwinds when pulled. From Ref. [11], reprinted with permission from AAAS. (c)-(e) are reproduced (adapted) from Ref. [12] with permission of The Royal Society of Chemistry. (f) Helices with decreasing cross-section height-to-width ratio. The structure moves from helix, to hemihelix, to hemihelix with multiple perversions. (g) Number of perversions as function of aspect ratio and prestrain. Adapted from Ref. [177] with permission.

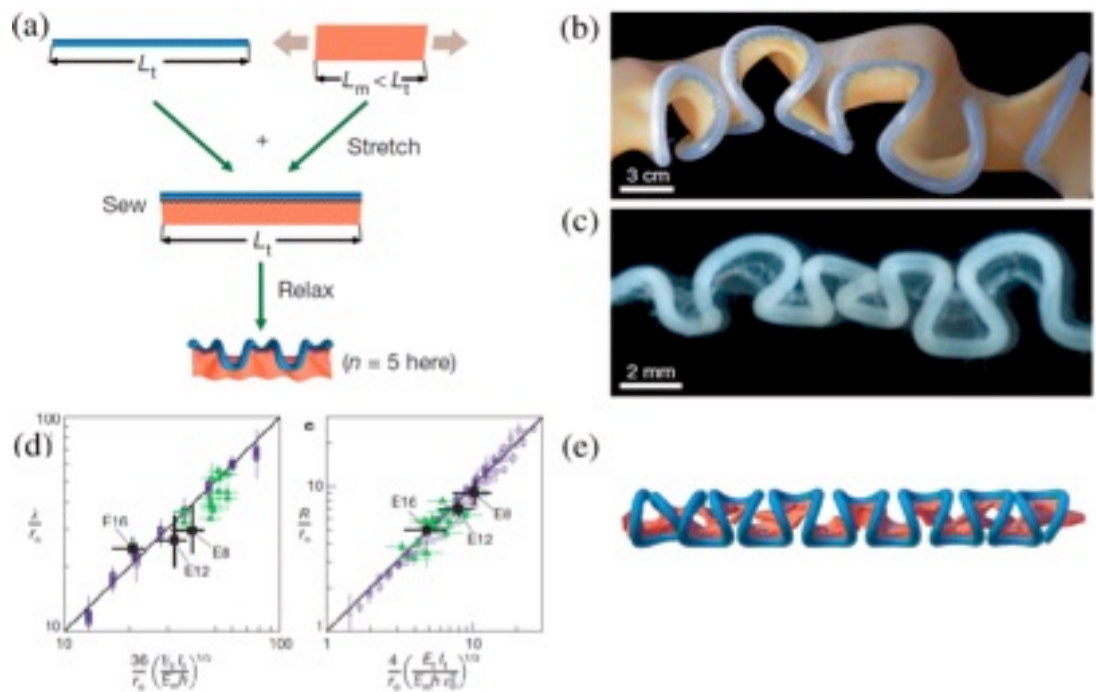

FIG. 30. (a-c) Thin stretched rubber sheets are stitched to unstretched rubber tubes to induce twisted patterns that resemble the looping guts in chick embryos in (c). (d-e) Scaling laws for the loop shape, size, and number at three stages in the development of chick guts. Reprinted by permission from Macmillan Publishers Ltd: Nature, Ref. [17] (Copyright 2011).

While many of the helical structures have a uniform chirality or handedness (i.e., are either left-handed or right-handed), change of handedness can occur under certain circumstances. For example, growing plant tendrils were found to switch handedness after the two ends were fixed in space; the tendrils would develop coexisting left-handed and right-handed parts connected by perversions (Fig. 29 a-b) [11,178,179]. Inspired by these tendril perversions, Huang et al. [12] bonded two elastomer strips of unequal length, one of which was prestretched, to create hemi-helices and performed finite element simulations to interpret their findings. Savin et al. [17] also constructed hemi-helical structures to mimic the looping guts in developing embryos. They prestretched a rubber sheet and stitch to it a rubber tube along the side (Fig. 30a). When relaxed, the tube developed a pattern featuring multiple hemi-helical segments connected by perversions (Fig. 30b), resembling the shape of chick guts (Fig. 30c).

Gerbode et al. [11] constructed bilayer silicone ribbons, one layer of which was prestretched, to study the physical mechanisms of the coiling and overwinding of the cucumber tendrils. Again, the topological constraint from the two fixed ends dictated that perversions would form, but the perversions that connect the helical segments of opposite handedness also allowed for rotations, causing an increase in the number of helical turns. Their study showed that when bending stiffness is smaller than twisting stiffness, the ribbon would unwind upon extension. However, for a helix with a round cross-section, the bending stiffness is always larger than the twisting stiffness, which indicates that overwinding would occur upon extension. As a result, the "twistless springs" can undergo axial extension simply through overwinding without paying an additional energy penalty for changing the curvature, a good strategy for creating soft springy tendrils that will stiffen upon further deformation [11].

Liu et al. [177] further investigated the transitions between a helix and a hemi-helical structure (with perversions). It was found that the twist buckling instability could prevent the system from going to the lowest energy state. These buckling modes have led to the formation of hemihelices that include multiple perversions but have higher energy than a simple helix. The system, once trapped in the metastable state, would have to be deformed by external forces and torques in order to return to the global energy minimum state. The researchers used a combination of experiments, finite element simulations, and theoretical analysis (based on Kirchhoff's rod theory) to demonstrate that the aspect ratio, which is closely related to the ratio between the bending stiffness and twisting stiffness, plays a key role in determining the number of perversions per unit length.

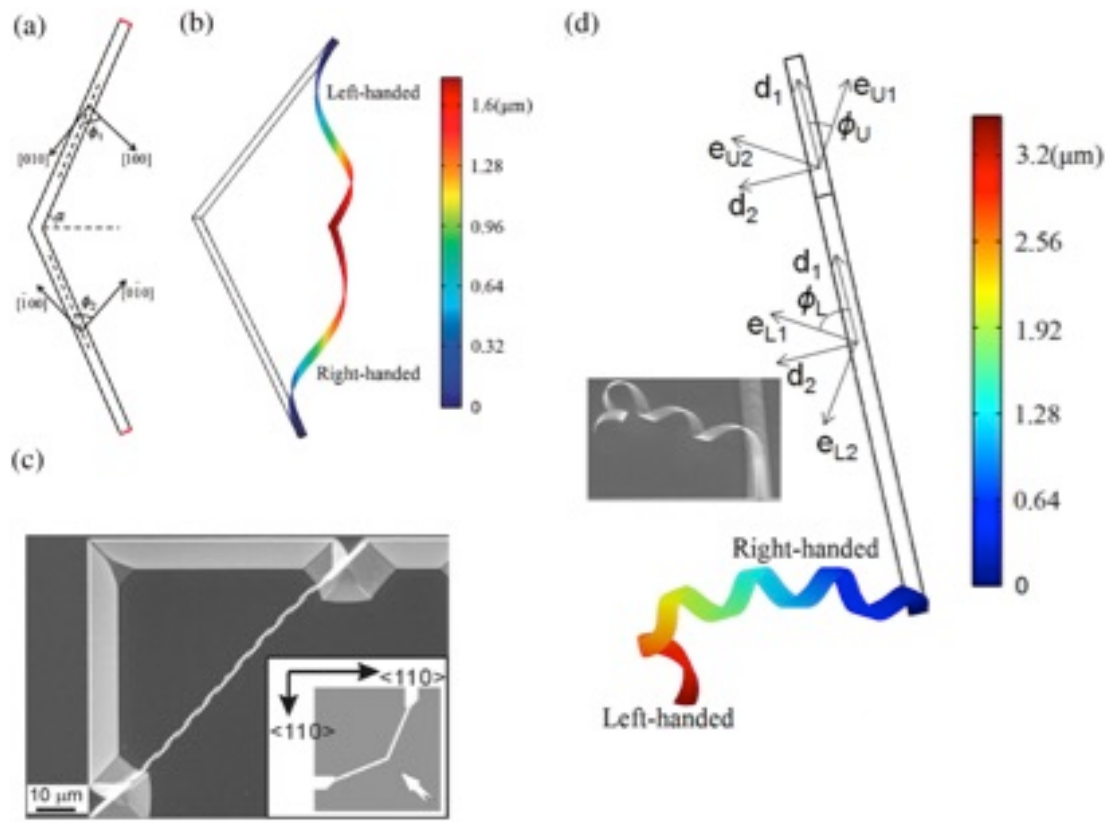

FIG. 31. (a) Geometry of a symmetric mesa design. Both ends (in red) are fixed. (b) A strained nanoribbon with symmetric left-handed and right handed segments (both ends being fixed). Here, $\emptyset_1 = \emptyset_2 = 75°$, and $\varepsilon_0 = 0.024$. (c) SEM image of a typical helical structure (diameter of 1.4 μm) formed by a V-shaped mesa with both ends fixed to the substrate. The two arms of the V-shaped mesa form helices, with opposite chirality. The inset shows the mesa design and the rolling direction of the helix as indicated with a white arrow. (d) A helical nanoribbon with both left-handed and right-handed segments with only one fixed end. Here, $\emptyset_L = 50°$, $\emptyset_U = 40°$ and $\varepsilon_0 = 0.05$. In the upper segment (of length 0.8 μm), the effective misfit strain tensor of the bottom layer is $\varepsilon_b = \varepsilon_0 e_{U2} \otimes e_{U2}$. In the lower segment (of length 2.4 μm), the effective misfit strain tensor of the bottom layer is $\varepsilon_b = \varepsilon_0 e_{L2} \otimes e_{L2}$. The color indicates the total displacement. (a), (b) and (d) are reproduced (adapted) from Ref. [169] with permission of The Royal Society of Chemistry. (c) and the inset of (d) are adapted from Ref. [138] (Copyright IOP Publishing, reproduced with permission, all rights reserved).

Yet another way of fabricating a helical ribbon with both left-handed and right-handed parts is by virtue of a V-shaped mesa design (Fig. 31a-c) [138,146]. For example, Fig 31 shows such a structure can be achieved by designing a mesa shape where $2\alpha + \emptyset_1 + \emptyset_2 = 270°$. In order to for the left- and right-handed helical segments to be symmetric, it should follow that $\emptyset_1 = \emptyset_2$. In Fig. 25b, it is set that $\emptyset_1 = \emptyset_2 = 75°$ and $\alpha = 60°$.

In all the cases discussed above, the changes in handedness occurred when the ribbon had two fixed ends. However, such geometric constraint is not a necessary condition for a switch in handedness. It was found that in SiGe/Si/Cr nanobelts left-handed and right-handed segments could co-exist when the misalignment angle was slightly larger than 45° and the tip had an influence towards the preferred chirality [138]. Although a qualitative interpretation of this behavior was given, quantitatively modeling of this phenomenon was not achieved until recently [169]. An FEM simulation approach was employed to model this behavior (Fig. 31b). A bilayer nanoribbon of length L = 3.2 μm, width w = 0.1 μm, thickness $h_1 = h_2 = 5nm$, and misfit strain $\varepsilon_0 = 0.05$ is partitioned into two connected regions of length 0.8 and 2.4 μm, respectively. The misalignment angle between the ribbon's long axis and the major bending axis <100> in the lower part and the upper part are both 50°. The resulting simulation shows co-existence of both left- and right-handed segments separated by a short perversion [11,178], in agreement with the experimental observations made by Zhang et al. [138] (Fig. 31d). Nevertheless, the physical mechanism of this perversion is different from that of tendril perversions since the boundary conditions are different.

## VI. Multi-stability in strained multilayer systems

Some mechanical structures can exhibit more than one stable shape [10,42,62,63,71,88]. For example, the lobes of the Venus flytrap (*Dionaea*) can be triggered to snap within a fraction of a second to capture insects [180]. Slap bracelets and tape springs are another set of multistable structures [13]. In recent years, multistable structures have received much attention because of their applications in micropumps, valves, deployable aerospace structures [10], mechanical memory cells [71], artificial muscles, bioinspired robots [19], and energy harvesting devices [43]. Such structures have inspired the design of deployable or smart actuation devices with multiple stable shapes, each of which can function independently.

## A. Multistability in Venus flytrap and bio-inspired structures

Forterre et al. [180] used the mechanical instability principle to interpret the snap-through of the leaves in the Venus flytrap. They put arrays of UV fluorescent markers on the surface of the leaves to calculate the principal curvatures ($\kappa_x$ and $\kappa_y$). In their experiment, an ultraviolet light was used to irradiate the Venus flytrap and a high-speed camera videotaped the trap closure process. From the recorded data, the researchers adopted the spatially averaged Gaussian curvature and mean curvature in order to simplify the analysis. Leaf-cutting experiments were also performed to measure the intrinsic principal curvatures in a closed trap, as shown in Fig. 32a. Moreover, a simple model based on elasticity theory was proposed which took into account the non-trivial coupling between bending and stretching of a plate. The leaf was modeled as a thin elastic shell with a radius L, thickness H, Young's modulus E, and intrinsic principal curvatures $\kappa_{x0}$ and $\kappa_{y0}$. It is assumed that at time t=0, $\kappa_x = \kappa_y = \kappa_{yo} = \kappa$ in order to simplify the analysis. The total elastic energy at time t is $U = U_{bending} + U_{stretching} = (K_x - K_{x0})^2 + (K_y - 1)^2 + \alpha(K_x K_y - 1)^2$, where $K_i = \kappa_i / \kappa$ (i = x, y) are the normalized principal curvatures, and $\kappa$ is the de-normalization factor for the principal curvatures. The dimensionless parameter $\alpha \sim L^4 \kappa^2 / H^2$ depicts the nonlinear coupling between bending and stretching.

By minimizing the total energy with respect to $K_x$ and $K_y$ (i.e., $\frac{\partial U}{\partial K_1} = \frac{\partial U}{\partial K_2} = 0$), at different values of $\alpha$, the leap shape can be resolved as a function of $K_{x0}$. It was then shown that the dimensionless parameter $\alpha$, which is related to the size, thickness, and curvature, controls bistability of the Venus flytrap (Fig. 32b). In particular, when $\alpha < \alpha_c \approx 0.8$, the system only has one minimum energy state; when $\alpha > \alpha_c \approx 0.8$, the system becomes multistable (Fig. 32b).

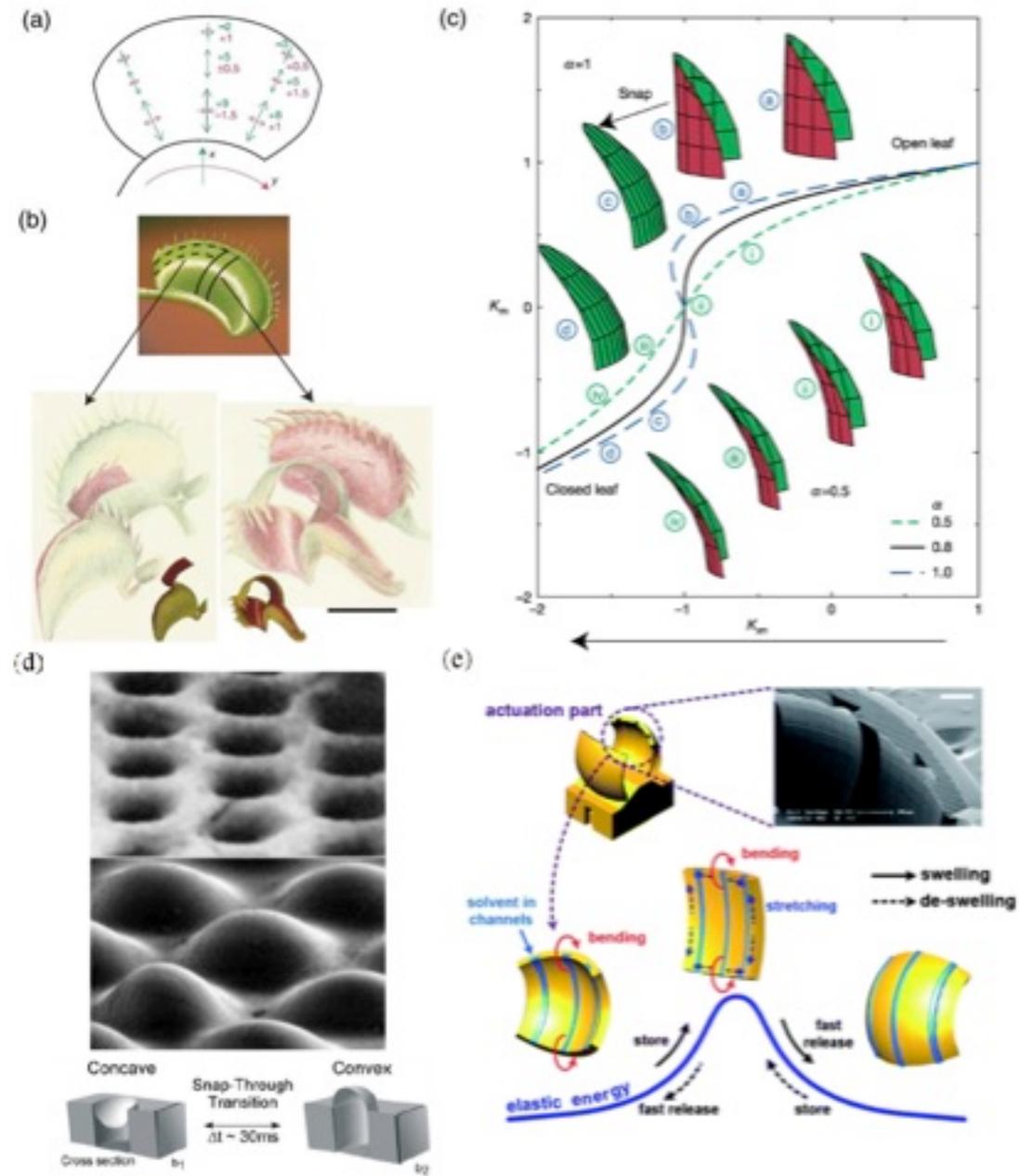

FIG. 32. (a) Measurement of the strain field; (b) the closed leaf is cut to illustrate the natural principal curvatures; (c) mean curvature as $K_m$ a function of $K_{xn}$. (a)-(c) are reprinted by permission from Macmillan Publishers Ltd: Nature, Ref. [180] (Copyright 2005). (d)

Snapping surface of concave microlenses that mimics the Venus flytrap. Adapted from Ref. [181] with permission. (e) A swelling-induced snapping microbot made of hydrogel shell with microfluidic channels embedded (a scanning electron microscope image embedded). Reproduced (adapted) from Ref. [182] with permission from The Royal Society of Chemistry.

Inspired by the Venus flytrap, Holmes and Crosby used a mechanical buckling mechanism to create bistable snapping surfaces [181]. Fig. 32d shows doubly-curved bistable shells generated through biaxial compression. PDMS patterned with holes is prestretched and crosslinked with a top layer of uncured PDMS. This forms concave and convex microlenses that exhibit snapthrough. Varying geometric parameters such as lens size and spacing can control the curvature of these lenses.

Lee et al. [182] further designed snapping robots using a similarly bioinspired strategy. A jumping microgel device was manufactured as shown in Fig. 32e. By creating a dome-shaped hydrogel shell that could swell or deswell in response to external stimuli and employing elastic instability, the researchers were able to achieve rapid actuation. The self-jumping device can accomplish a snap-through motion within a period of 12 ms. Upon swelling, the microgel legs of the device snap-buckle, resulting in a rapid jump. The power density of this device approaches that of human muscle.

Shankar et al. [183] show that a lightweight snap-through actuator can be designed through the use of azobenzene-functionalized polymers. By using bistable arches, the authors were able to generate a snap through with a speed as high as 100mm/s. These actuators buckle in response to light, making them especially suitable as small remotely and precisely triggerable actuators. Leong et al. [184] have produced a similar snapping micro-gripper structure that can be remotely activated both by chemicals and heat. These grippers can be used to perform cellular-scale tasks on command.

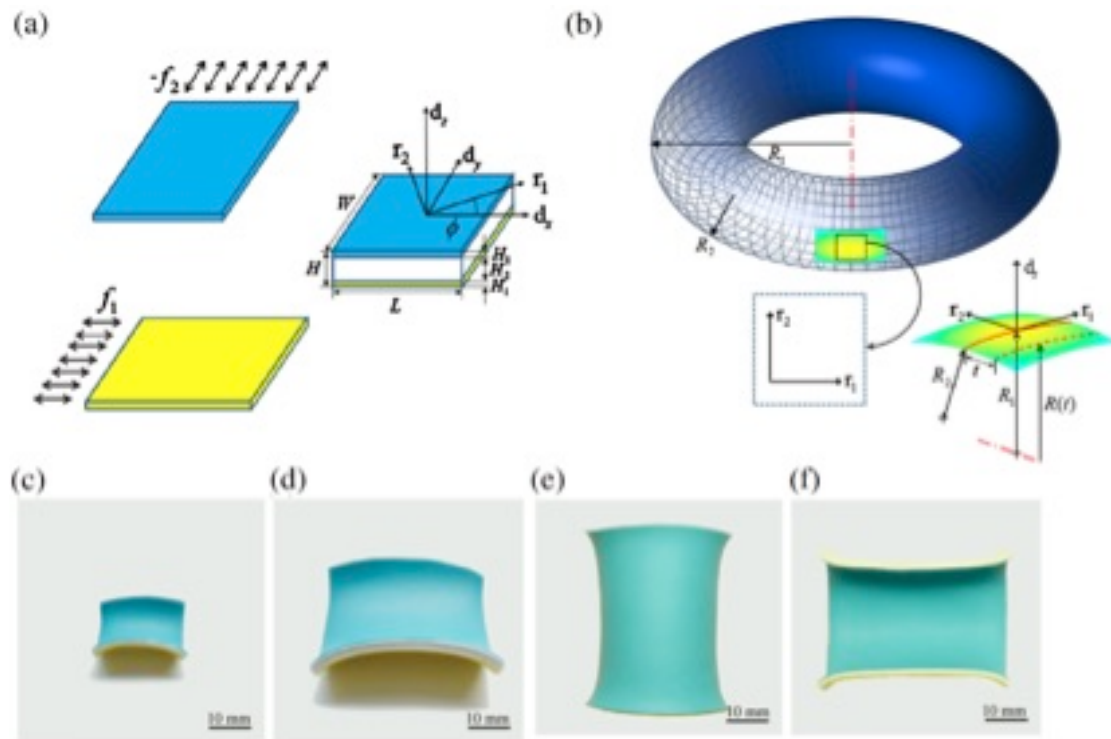

FIG. 33. (a) Blue and yellow latex sheets were perpendicularly prestretched and bonded to a thicker elastic strip. The multilayer sheet forms a doubly curved shape that conforms to a torus (b). Small (c) and large (d) thin squares form saddles, while thin strips form stable semi-cylinders. (e,f). Reprinted with permission from Ref. [185] (Copyright 2012 by the American Physical Society).

## B. Bistable and neutral stable shells

Chen et al. [185] developed a theoretical model to study bistability in strained multilayer systems complemented with tabletop experiments. Monostable and bistable shapes were manufactured by prestretching two pieces of rubber sheets uniaxially along perpendicular directions, and then bonded with layer of adhesive elastic sheet made of acrylic (Fig. 33a, c-f). The composite layers can either be monostable or bistable depending on the geometric dimensions: they are monostable when the system is either narrow or thick but bistable otherwise. The geometric and mechanical conditions of bistability were studied through a theoretical model that models the deformation of an initially flat elastic strip onto the surface of a torus (Fig. 33b) with two geometric parameters ($\kappa_1$ and $\kappa_2$). By comparing the bending and in-plane stretching energy in the model, two dimensionless parameters (the first related to both the mechanical driving forces and geometric parameters, and the second purely associated with the forces) were identified to be controlling bistability. The first parameter, $\xi = \sqrt{f/EH} W/H$, involves the surface stress f, the Young's modulus E, the width W, and the thickness H. When it is below the threshold value, the multilayer system is monostable, as shown in Fig. 34c and 34d; when it goes beyond the threshold; the system becomes bistable (Fig. 34e-f). Noticeably, the parameter $\xi$ is essentially equivalent to the geometric parameter $\eta \equiv W\sqrt{\kappa/H}$ [15,149,185], but the former does not involve the unknown curvature $\kappa$, therefore is more suitable for the purpose of designing multistable structures. Guo et al. [22] employed this model to address the shape transition and associated change in multi-stability as discussed before.

The second parameter, $\beta \equiv f_2/f_1$, dictates the nature of the multistability. $\beta < 0$ is a necessary but not a sufficient condition for bistability (Fig. 34a, e-g); when $\beta \equiv 1$, the system can exhibit neutral stability (Fig. 34c, h) if the first parameter $\xi$ is above the critical value.

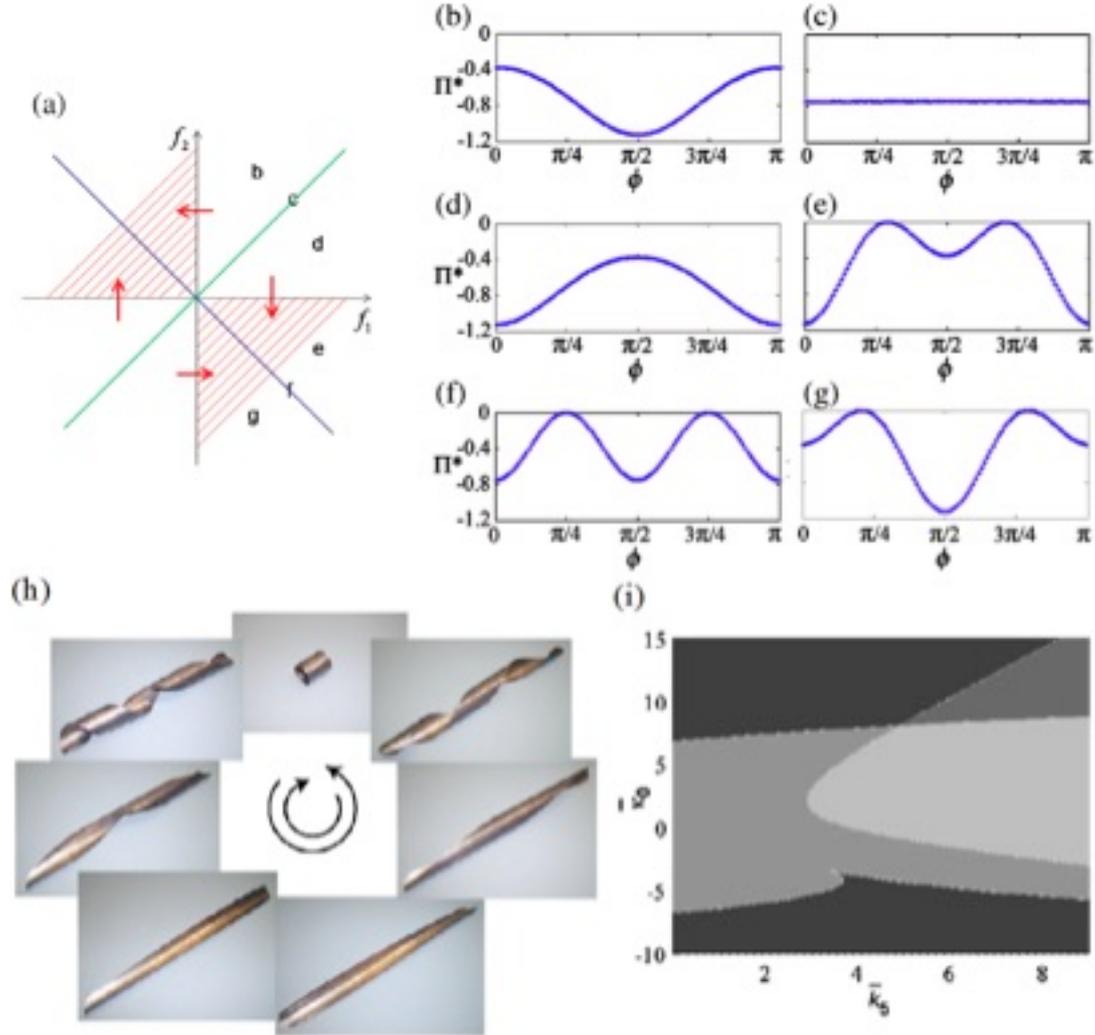

FIG. 34. (a) Multistable shell design space, with bistability in red. (b-g) Total strain energy as a function of misfit axis orientation, for orientations: $f_2/f_1 = \sqrt{3}$, $f_2/f_1 = 1$, $f_2/f_1 = 1/\sqrt{3}$, $f_2/f_1 = -1/\sqrt{3}$, $f_2/f_1 = -1$, and $f_2/f_1 = -\sqrt{3}$. (a)-(g) are reprinted with permission from Ref. [185] (Copyright 2012 by the American Physical Society). (h) A variety of shapes of a pre-stressed shell structure with zero-stiffness. The shell can be transformed from one configuration to another either along the clockwise or anti-clockwise path and each shape can be held merely by friction with the table's surface. First published in Journal of Mechanics of Materials and Structure in 2011, from Ref. [186] (published by Mathematical Sciences Publishers copyright 2011). (i). Adapted from Ref. [187] with permission (Copyright 2011 ASME).

In fact, structures featuring neutral stability have been created as zero-stiffness elastic structures [16,186,187]. Guest and co-workers studied a prestressed shell structure that can deform with zero twist rigidity as shown in Fig. 34h. Experiments and analytical modeling have been carried out to gain insights on such novel mechanical properties [186,187]. Guest et al. [186] proposed a simple analytical model based on the assumption that the shell is not extensional and that the curvature is uniform throughout the shell (neglecting possible edge effects). As a result, the shell's mid-plane can be considered to lie on the surface of a cylinder and bending along any arbitrary axis is energetically equally favorable. Seffen and Guest [187] modified these governing equations to obtain analytical solutions for both the opposites-sense and same-sense prestressed shells that exhibit bistability and neutral stability, respectively. For isotropic materials, previous studies showed that an elastic shell enters a neutrally stable state (with zero twist rigidity) under same-sense prestressing when the prestress is sufficiently large, which was observed in heated plates [188] although it had not been quantitatively characterized as in a neutral stable state until recently [186,187]. By contrast, opposite-sense prestressing can result in a bistable state when the driving force is large enough, which is similar to a bifurcation phenomenon first studied by Hyer that occurred during the curing of unsymmetrical laminates. Hyer discovered that a flat plate with anisotropic mechanical properties deformed initially into a saddle shape with a negative Gauss curvature, but when the curing temperature further increased, the shell could no longer be stable in a doubly curved shape and would bifurcate into a configuration with nearly zero Gauss curvature, albeit with two opposite bending directions [188].

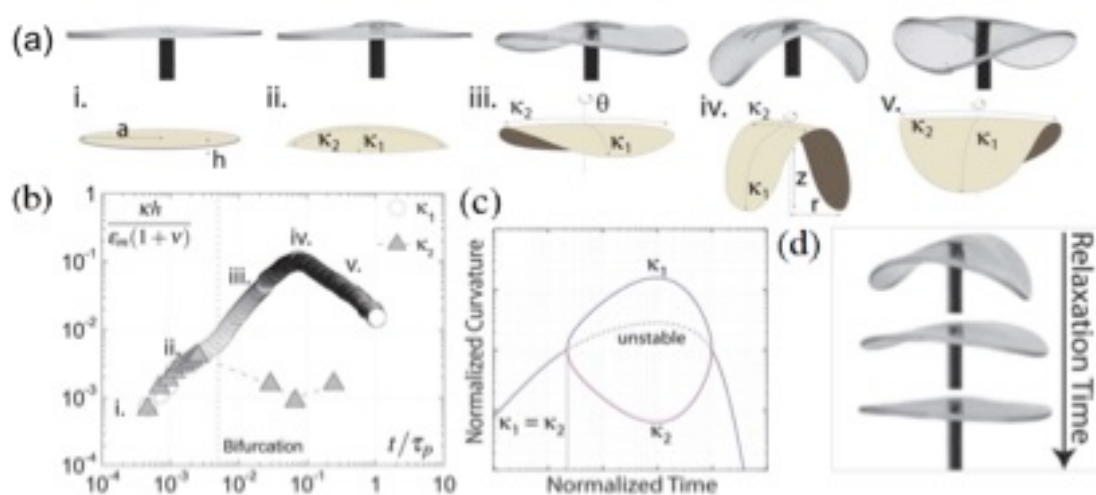

FIG. 35. (a) Images and schematic diagrams of a circular disc buckles axisymetrically with principal curvatures $\kappa_1 = \kappa_2$, and bifurcates with two distinct positive curvatures. (b) Normalized curvature versus non-homogeneously swollen time. (c) By minimizing the total strain energy, the relationship between normalized curvature and normalized time is obtained and is in good agreement with experimental data. (d) The circular disc relaxes back to its original flat state. Reproduced (adapted) from Ref. [16] with permission of The Royal Society of Chemistry.

The mechanisms for such bifurcation phenomena have since been investigated by a number of researchers [15,16,185,187-190]. Holmes et al. [16] investigated the bending of slender crosslinked polydimethylsiloxane beams by soaking one side of the beams in one of two different solvents (analogous to applying a thermal gradient) and measured the resulting curvature over time. They then developed analytic solutions that provided quantitative characterization of curvature timescales and normalized parameters. It was found that the beams bend quickly to their maximum curvature but relax slowly. The author further investigated axial symmetric quasi-2D discs characterized by two principal curvatures. When bifurcation occurs, Karman plate theory predicts that the two principle curvatures become unequal in magnitude and rotate azimuthally, while the material points stay horizontally locked. This phenomenon, analogous to thermal swelling, was experimentally verified by marking points on the plates - during dynamic rotation, the points moved only in the vertical direction.

Moreover, researchers have designed and studied new devices consisting of multiple pre-stressed pieces to exhibit multistability [191,192]. For example, Lachenal et al. [191] created a joined structure from two pre-stressed flanges to have two stable shapes and analyzed the mechanical behaviors using a combination of experiments, finite element analysis, and a simple analytical model. Pirrera et al. [192] further developed cylindrical lattices comprised of hinged helical strips to exhibit bistability and neutral stability that mimic the multistable behaviors of the virus bacteriophage T4. These multistable structures can serve as outstanding candidates for energy absorption devices, morphing structures, and deployable structures for a variety of engineering applications.

## VII. Outlook

This review provides a comprehensive overview of state-of-the-art techniques for mechanical self-assembly of strain-engineered flexible layers. By programming prescribed strain into these flexible layers, they can be made to wrinkle, roll, or twist predictably both at manufacturing and when activated by stimuli. We explore the mechanics of strain-induced bending multilayers as well as of spontaneous helix formation, and show experimental examples manufactured from materials such as hydrogel bilayers and liquid crystal polymer networks. Shape transitions and multistability in the resulting structures, including in helical ribbons, are also discussed. In order to limit the article's length, certain aspects of large soft layers deformation like creasing instability [187,188] are not discussed. Various methods to harness this kind of nonlinear behavior was shown in a recent review [193].

While there have been significant advances in techniques to manufacture self-assembling strain-engineered layers, gaps remain in our understanding of folding mechanics, and large improvements in the mechanical properties of most layers are necessary before they are commercially viable. Shape morphing layers are also slower and weaker than other classes of actuators. Although piezoelectric materials can actuate on the order of milliseconds, composites that rely on changes in temperature, humidity, or light take minutes or hours to fully morph from one shape to another. Certain voltage stimulated polymer composites can lift many times their own weight, but most layers can lift only a small fraction. This is a result of both the relative slowness of environmental signals and their very low energy density. In addition, some layers have significant issues with hysteresis, which limits their ability to be reused [194-196]. More work needs to be done both in improving environment-triggered layers for use in adaptive surfaces and in developing flexible materials that can actuate in response to more energy-dense chemical or electrical triggers.

Shape-morphing flexible layers have a wide range of applications as sensors, actuators, and adaptive surfaces. Because they passively actuate in response to environmental changes like temperature, humidity, salinity, and pH, these materials and structures make excellent candidates for environment sensors. Unlike traditional passive sensors, flexible layers can adopt any shape and thus it is significantly easier to add them to structures and devices. In addition, their thin profile and potential sensitivity makes them suitable for artificial skin, as they are simpler than the matrices of hard sensors currently used in the field. Their small footprint and low material cost allow them to be constructed cheaply and efficiently.

One can also imagine broad uses as passive actuators, such as blinds that open and close in response to temperature and sunlight or solar panel actuators that mimic plants to track the movement of the sun. Passive solar panel movement increases the overall efficiency of the panels, which can add up to a huge amount of energy over large installations. Briefly discussed in the review are layers that exhibit multistability over a single stimulus. This behavior, if harnessed through clever engineering as actuation for an origami robot, could allow the robot to function indefinitely without an onboard power source. This is important primarily for scientific robots operating in extreme environments such as the deep sea, artic tundra, and other planets. If the robot can passively or semi-passively move autonomously, it frees up power for scientific instruments. More broadly, self-assembling layers that transition under stimuli are useful for all systems where functionality is to be built into form.

## VIII. Acknowledgment

This work is supported by Society in Science-Branco Weiss fellowship administered by ETH Zürich (Z.C.), the Natural Science Foundation of China (Nos. 51322201 and 51475093) and Science and Technology Commission of Shanghai Municipality (No. 12520706300).